\documentclass[twocolumn]{emulateapj}

\newcommand{\kms}{km~s$^{-1}$}

\newcommand{\etal}{{\it et al.}}
\newcommand{\ie}{{\it i.e.}}
\newcommand{\eg}{{\it e.g.}}
\newcommand{\be}{\begin{equation}}
\newcommand{\ee}{\end{equation}}

\newcommand{\ra}{$R_A$}

\newcommand{\kmsMpc}{km~s$^{-1}$ Mpc$^{-1}$}

\newcommand{\ins}{ {\bf in}}
\newcommand{\inp}{{\bf in+}}

\shorttitle{SFI++ TF Template}
\shortauthors{Masters \etal}

\begin{document}

\title{SFI++ I: A New I-band Tully-Fisher Template, the Cluster Peculiar Velocity Dispersion and $H_0$.}
\author{Karen L. Masters\altaffilmark{1,2}, Christopher M. Springob\altaffilmark{1,3}, Martha P. Haynes\altaffilmark{1,4} \& Riccardo Giovanelli\altaffilmark{1,4}}
\altaffiltext{1}{Center for Radiophysics and Space Research, Cornell University, Ithaca, NY 14853, haynes@astro.cornell.edu, riccardo@astro.cornell.edu}
\altaffiltext{2}{Harvard-Smithsonian Center for Astrophysics, 60 Garden Street, Cambridge, MA 02138,kmasters@cfa.harvard.edu}
\altaffiltext{3}{Naval Research Laboratory, Remote Sensing Division Code 7213, 4555 Overlook Avenue, SW, Washington, D.C., 20375, Christopher.Springob@nrl.navy.mil}
\altaffiltext{4}{National Astronomy and Ionosphere Center, Cornell University, Space Sciences Building, Ithaca, NY 14853. The National Astronomy and Ionosphere Center is operated by Cornell University under a cooperative agreement with the National Science Foundation.}

\begin{abstract}
The SFI++ consists of $\sim$5000 spiral galaxies which have measurements suitable for the application of the I-band Tully-Fisher (TF) relation. This sample builds on the SCI and SFI samples published in the 1990s but includes significant amounts of new data as well as improved methods for parameter determination. We derive a new I-band TF relation from a subset of this sample which consists of 807 galaxies in the fields of 31 nearby clusters and groups. This sample constitutes the largest ever available for the calibration of the TF template and extends the range of line-widths over which the template is reliably measured. Careful accounting is made of observational and sample biases such as incompleteness, finite cluster size, galaxy morphology and environment. We find evidence for a type-dependent TF slope which is shallower for early type than for late type spirals. The line-of-sight cluster peculiar velocity dispersion is measured for the sample of 31 clusters. This value is directly related to the spectrum of initial density fluctuations and thus provides an independent verification of the best fit WMAP cosmology and an estimate of $\Omega^{0.6} \sigma_8 = 0.52\pm0.06$. We also provide an independent measure of the TF zeropoint using 17 galaxies in the SFI++ sample for which Cepheid distances are available. In combination with the ``basket of clusters'' template relation these calibrator galaxies provide a measure of $H_0 = 74\pm2$ (random) $\pm6$ (systematic) \kmsMpc.

\end{abstract}

\keywords{cosmological parameters --- distance scale --- galaxies: clusters: general --- galaxies: distances and redshifts --- galaxies: fundamental parameters}

\section{Introduction}
The Tully-Fisher relation \citep{TF77} is an empirical relationship between the rotational velocity and absolute magnitude (or luminosity) of spiral galaxies. The application of this relation to measure distances to spiral galaxies has had an enormous impact on extragalactic astronomy and cosmology since its discovery. The Tully-Fisher (TF) relation for spirals (along with the $D_n-\sigma$ relation for ellipticals) dramatically increased the volume in which redshift-independent distances could be measured. The TF relation and its scatter can also be used to constrain models of disk galaxy formation, and by studying the relations variation with environment, galaxy type and its evolution over the age of the universe many clues into how the disks of galaxies are assembled have been found. A reliable template relation is an essential starting point to all this, without which no trust can be placed in other conclusions. Like all empirically determined scaling relations, the selection effects and make-up of a TF sample can introduce non-trivial biases into the derived template which if not taken care of could introduce spurious results. With improved statistics and parameter measurement both in galaxy surveys and simulations the need for an accurate, unbiased template grows even stronger.

The TF relation has historically played an important role in cosmological parameter determination, in particular by allowing measurements of $H_0$ to be extended out of the very local universe. In fact a review of TF measurements of $H_0$ over time shows that the technique has done surprisingly well. 
The value of $H_0$ from TF studies played a significant role in the debate between high and low values of $H_0$, with cluster peculiar velocities, the Malmquist bias and even dust extinction intervening to provide a range of measures of $H_0$ from TF. The Hubble Key Project was initiated to resolve this debate and found $H_0=71\pm2\pm7$ \kmsMpc ~from a combination of various TF surveys and Cepheid data \citep{S00}. Similarly \citet{G97a} found $H_0=69\pm5$ \kmsMpc ~from a combination of the I-band TF relations in 24 clusters and a small set of Cepheid calibrators; and \citet{TP00} found $H_0=77\pm8$ \kmsMpc ~from their sample of 5 clusters. In Section 7 below we derive $H_0 = 74\pm2\pm6$ \kmsMpc ~from the sample discussed in this paper. Even in todays era of ``precision cosmology'' the TF constraints on $H_0$ have similar random error to the best available measurements combining WMAP and large scale structure data (\eg ~$H_0 = 74\pm2$ \kmsMpc ~from a combination of WMAP and 2dFRGS; \citealt{Sa06}), and better than WMAP alone which measures $H_0 = 73\pm3$ \kmsMpc ~(with the assumption that the universe is flat; \citealt{Sp06}). 

 \citet{TF77} first calibrated the TF relationship in the B band using a sample consisting of 10 galaxies with $D<8$ Mpc and eight galaxies in the Virgo cluster. They measured $M - 5 \log h = -19.13 - 6.25 (\log W - 2.5)$, where $M$ is the absolute magnitude, $W = 2 v_{\rm max}$ is the width of the Doppler broadened spectral line and $h = H_0/100$ \kmsMpc. This is equivalent to  $L \propto v_{\rm max}^{2.5}$. In the B band, extinction (both internal to the galaxies and from our Galaxy) plays an important role in adding to the scatter of the TF relation.
\citet{A79} argued that the relationship in the NIR would have a smaller scatter. They observed that the slope steepened in the H band, more nearly approximating the expected $L \propto v_{\rm max}^4$ from simple physical arguments. Much of the most successful early work using TF used the \citet{A80} calibration based on the distance to M31 and M33, and the assumption that $L \propto v_{\rm max}^4$. The uncertainty in these (and other) early TF templates from small samples of nearby calibrator galaxies was obviously dominated by small number statistics and since the samples did not cover a wide dynamic range of galaxy masses they also provided little leverage on the measurement of the TF slope. Several immediate attempts were made to extend the calibration, both by looking at cluster samples (where all galaxies could be assumed to be at roughly the same distance) and obtaining more local calibrators. In the late 1980s it was argued that the I-band TF relation could improve the scatter in the relation further \citep{BM87}. This band, intermediate between B and H, retained most of the advantages gained by moving into the NIR but had the additional advantage of allowing CCD photometry.

 It should be obvious that using a strictly magnitude-limited sample to calibrate the TF relation will result in an underestimate of the slope, as at the low-width end only galaxies scattered above the line will be included. An early solution to this potentially strong bias was the adoption of so-called {\it inverse} fits in which the magnitude was treated as the independent variable. Such a fit would not suffer from incompleteness bias in the simple case where selection depended only on absolute magnitude. Unfortunately, realistic samples do not have such simple selection criteria, and in particular internal extinction corrections, which have been shown to depend on galaxy luminosity, and therefore implicitly on the rotation width \citep{G94,M03} make bias corrections more complicated. Arguments also exist in the literature that cluster samples constitute {\it complete} or {\it volume-limited} samples and therefore the TF relation derived from such a sample will not suffer from incompleteness bias. As first discussed in \citet{S95} this is not the case. An explicit or implicit magnitude limit in a cluster sample will have a the same impact on the measured TF relation as a sample with a full spread of distances. 

 A biased TF template slope can introduce several subtle, {\it qualitative} biases into many of the conclusions which might be drawn from its application. For example Malmquist bias-like effects mean that the more distant objects in any given sample are more likely to be at the large-width end of the relation while the most nearby objects are more likely to be at the low-width end of the sample. If the template is biased shallow, this means that distant objects will preferentially have spurious negative magnitude offsets (\ie ~they appear to be artificially brighter than the biased template), while nearby objects will preferentially have spurious positive offsets. If these offsets were interpreted as being due to the peculiar velocities of the galaxies, this would produce a spurious infall region in the local universe and excess expansion in the more distant parts of the sample.  Determinations of the morphological dependence of the TF relation can be impacted by bias on the slope too. Early type spirals (Sas) are preferentially brighter and have larger rotation widths than later type spirals. A TF template biased shallow might then lead to the conclusion that the zeropoint for earlier type spirals was brighter than that for late type spirals. In a high redshift sample which will only have galaxies at the large width end of the local relation, comparison with such a biased local relation could easily lead to a spurious determination that the zeropoint of the TF relation brightens with redshift.

 In fact most high redshift studies of the TF relation use the relatively small samples of \citet{PT92} or \citet{MFB} as a $z=0$ comparison. The \citet{PT92} B, R and I-band calibration is based on the TF relation of only 6 local spirals for which Cepheid distances were available, in combination with a small sample ($\sim 30$ galaxies) from the Ursa Major cluster. \citet{MFB} use a sample of only 14 galaxies in the Fornax cluster to provide an I-band calibration. No attempt is made to account for sample selection biases in either sample. As discussed in \citet{MK06}, larger and better calibrated local samples already exist. The \citet{TP00} calibration uses a significantly larger sample (155 galaxies in 5 clusters at I-band; 91 galaxies at B and R and 65 at B-band), however the work still makes no attempt to account for selection biases relying on the erroneous idea that cluster samples are complete. \citet{KF02} use a sample of 196 nearby field galaxies but just fit the {\it inverse} U, B and R-band relations in an attempt to account for bias, and make no attempt to account for the effect of peculiar velocities which also have the potential to bias the template slope. No fully bias corrected local template exists in optical wavelengths.

The first comprehensive discussion of the effect of incompleteness bias on the TF template relation came from \citet{W94}. Here it was explicitly demonstrated that the {\it inverse} TF relation is not bias free. A method was introduced which used Monte Carlo simulations to calculate the bias based on the assumption of a global linear TF relation, a fixed scatter and a single magnitude limit. Giovanelli \etal~ (1997b; hereafter G79b) extended this technique to account for a TF scatter which varied with line width, and to provide realistic accounting of sample selection criteria. They used this to calibrate I-band TF relation with a sample of 555 spiral galaxies in the fields of 24 clusters. They found $M - 5 \log h = -21.01 - 7.68 (\log W - 2.5)$, or $L \propto v^{3.1}$. They observed that the scatter in the Tully-Fisher relation varies with the rotational velocity of the galaxy, from about 0.4 mag at the low width end, to less than 0.3 mag at the high width end, resulting in distance errors for individual galaxies of 15--20\%.

 Much of this paper is devoted to the derivation of a new template TF relation in the I-band. We follow a method similar to that used in G97b and use a sample of galaxies which builds on that cluster sample, but has a significant amount of new data (including the addition of several new clusters) making a total sample of 807 galaxies in the fields of 31 clusters. This sample constitutes the largest ever available for the calibration of the TF relation, and extends the range of the calibration between $v_{\rm max}=$ 50--360 \kms ~($\log W =$ 2.0--2.9). In combination with the much larger field sample (Springob \etal ~2007) this work will finally allow for studies of the extent, non-linearity and morphological dependence of the TF relation at high statistical significance. 

This new I-band template is warranted both because of the addition of significant amounts of new data to the sample and because of changes in corrections to the raw data. In particular, new corrections for rotation velocities derived from observations of both the 21cm and H$\alpha$ lines of neutral hydrogen (HI and H$\alpha$) are expected to have an impact on the derived template. New simulations have been performed to study the instrumental effects on measuring HI widths \citep{SH05}, and much greater care has been taken in combining HI and H$\alpha$ rotation measures (Catinella, Haynes \& Giovanelli 2006). The addition of significant amount of new data also alters the completeness and morphological distribution of the sample, which must be taken into account in rederiving the template. 

In Section 2 we discuss the observables needed for TF. Section 3 covers the sample selection and bias corrections. In Section 4 we discuss individual cluster TF and the impact of environment. Section 5 combines the individual cluster templates into a global template. In the process we measure the peculiar velocity of each cluster, from which a cluster peculiar velocity dispersion is calculated. This quantity depends relatively simply on the power spectrum of initial density fluctuations and is used to estimate $\Omega^{0.6} \sigma_8 = 0.52\pm0.06$. In Section 6 we describe the characteristics of the scatter of the TF template. Section 7 discuss an alternate zero-point calibration which uses Cepheid distances to 17 galaxies in the SFI++. In combination with the ``basket of clusters'' zeropoint this gives a measure of $H_0 = 74\pm2\pm6$ \kmsMpc. The final section of the paper provides a summary of the conclusions. In future work including Masters \etal ~(2007, in prep.) and Springob \etal ~(2007; also see \citealt{KLM}; \citealt{CS}) this new template will be applied to a larger sample of spiral galaxies in the field, from which the peculiar velocity field, and therefore the mass distribution, of the local universe will be studied. 

\section{Tully-Fisher Observables}
The sample discussed here will be referred to as the SFI++. This sample builds on the all-sky SFI (Spiral Field I band) and SCI (Spiral Cluster I band) samples discussed in a series of papers in the 1990s \citep{G94,G95,G97a,G97b,H99a, H99b}, but also includes data from the SC2 sample \citep{DD,D99}, the theses of \citet{NV} and \citet{BC} and the HI archive presented in \citet{SH05}. The entire SFI++ sample will be presented in Springob \etal ~(2007). Subsets of this data have also been presented in the theses of \citet{KLM}, \citet{KS} and \citet{CS}. The bulk of the southern hemisphere data is presented in \citet{MFB}. The full SFI++ contains $\sim$5000 spiral galaxies with observations suitable for deriving TF distances. In this paper we consider a subset of that data consisting of galaxies in the vicinity of 31 nearby clusters. This we will refer to as the ``template sample" of the SFI++. Most clusters in this sample were specifically targeted for I-band observations either as part of \citet{G97a}, \citet{D99} or \citet{V04}. Fields were chosen to include at least one spiral galaxy for which dynamical information (either a HI width or optical rotation curve) was available at the time of observation, and the centers of the fields were adjusted to include as many other good TF candidates (inclined, undisturbed spirals) as possible which might later be targeted for spectroscopy. The sample is therefore in no sense magnitude or angular diameter limited, but unlike SFI/SCI does include all types of spirals.

\subsection{Optical Imaging Data}
All optical data discussed here is I-band imaging data taken by members of the Cornell Extragalactic group and their collaborators at various telescopes. Data for the SFI sample of galaxies is presented in \citet{H99b}. This was a combination of images in the Northern hemisphere taken for various projects, and the similar \citet{MFB} survey in the Southern sky from which data were carefully recalibrated to make a homogeneous sample. For the SFI++ sample discussed here new data were taken in about 1900 fields north of $\delta = -2$\arcdeg ~using the 0.9m KPNO telescope. The fields covered various clusters and TF candidate galaxies and were positioned in an attempt to optimize the number of TF candidate galaxies observed. These imaging data were intended to be complete to a diameter limit of about 0.5 arcmin in I band, and more than double the number of TF candidate galaxies for which I-band imaging data is available. 

 The I-band images were reduced as described in \citet{H99b} to provide the observed total apparent magnitude and axial ratio of the galaxy. The observed axial ratio, $a/b$ of the galaxy comes from isophote fitting and is corrected for seeing as in G97b.
All magnitudes are corrected Galactic extinction using the DIRBE dust maps \citep{SFD98}, including those magnitudes from the SFI and MFB imaging data which in previous publications used the \citet{BH78} correction.  Extinction due to dust internal to the galaxy itself is accounted for using the extinction correction derived in \citet{G94}. A small type-dependent k-correction as discussed in \citet{H92} is also applied.

\subsection{Rotation Widths}
Rotation widths for the SFI++ sample come from a mixture of 21cm line (HI) global profiles and optical rotation curves (ORCs). Where they are available and of good quality, HI global profile widths are used in preference to ORCs. The HI extent of a normal galaxy is typically about twice its optical size, so HI global profiles are capable of tracing the rotation of galaxies to larger radii, and also have the advantage of being independent of the assumed position angle of the galaxy. Approximately 60\% of the galaxies in the SFI++ have their rotational velocities measured using HI, the other 40\% use ORCs. Just less than 20\% have measurements available from both methods. A comparison of widths from the different methods is presented in Catinella \etal ~(2007).

\subsubsection{HI Spectroscopy} \label{HI}
The 21cm HI spectroscopy data for the SFI++ is a combination of data published in \citet{H99a} and \citet{SH05}, reprocessed by \citet{SH05}. For the purposes of the TF relation we are interested only in the systematic velocities and velocity widths, as discussed in Section 3.2 of \citet{SH05}. Raw widths are measured using an algorithm that fits low order polynomials to either side of the HI profile and then calculates the width at 50\% of $f_p - rms$ (the peak flux minus the $rms$) on either side of the HI profile. These raw widths then have to be corrected for various instrumental effects, $\Delta_s$, the impact of turbulent motions $\Delta_t$, the disk inclination, and also relativistic effects from the observed redshift which broaden the observed profile by a factor $(1+z)$. 
Where the HI widths used here differ from those in \citet{H99a} and earlier work of this group, is in the magnitude and trend with $S/N$ of the instrumental corrections, $\Delta_s$. \citet{SH05} present the results of simulations to study the impact of spectrometer resolutions, smoothing and $S/N$ on the measured widths, and their prescribed corrections have been applied to {\bf all} the HI widths used here (including the reprocessing of older widths by CS). This width correction $\Delta_s$ is always a small positive number, which in \citet{SH05} increases linearly with $\log (S/N)$ between $S/N =$ 4--12.5. The previous correction for instrumental broadening decreased as the $S/N$ ratio increased. For galaxies in a single cluster sample where $S/N$ roughly correlates with rotation width this new correction should then tend to shallow the TF slope slightly relative to that found with the old correction in G97b. The correction for turbulent motions is also done in a slightly different way from previous work. \citet{SH05} discuss the merits of simple linear subtraction as opposed to a subtraction in quadrature. $\Delta_t = 6.5$ \kms ~is assumed.

\subsubsection{Optical Spectroscopy}

Optical rotation curves are used to fill in areas of the sky or galaxies not accessible to 21cm HI observations. For example about half of the rotation widths in the southern hemisphere are from H$\alpha$/[NII] RCs taken at the 2.3m telescope at the Siding Springs Observatory. There are also a significant number of ORCs taken using the long slit spectrograph on the Hale 5m telescope at the Palomar Observatory (see \eg ~Catinella, Haynes \& Giovanelli 2005). As discussed in \citet{BC} and \citet{C05} there are several algorithms which can be used to measure rotational velocities from ORCs. All of the widths used in the SFI++ have been derived by fitting a function to the folded ORC. The function used, the ``polyex'' model \citep{GH02,BC,C05} has the form
\be
\label{polyex}
V_{\rm PE}(r) = V_0(1 - e^{-r/r_{\rm PE}})(1+\alpha r/r_{\rm PE})
\ee
where $V_0$ gives the amplitude of the ORC, $r_{\rm PE}$ gives the exponential scale of the inner rise of the ORC, and $\alpha$ gives the slope of the outer part of the ORC. This function has no physical meaning, but provides a useful empirical fit to a wide variety of ORC shapes. Widths are derived from this fit by taking its value at the location of the optical radius, $r_{\rm opt}$ which is the radius enclosing 83\% of the optical light (here derived from I-band imaging). This raw width is then corrected for inclination and cosmological broadening in the same way as the HI widths.

As shown in \citet{C07} there are systematic trends of the difference between widths measured from ORCs and HI widths. HI widths are systematically larger than ORC widths presumably because HI disks are on average twice as large as H$\alpha$ disks and ORCs in general are flat or still rising beyond the optical radius \citep{C06}. \citet{C07} also show that there is a slight systematic trend of this difference with the extent of the ORC, in the sense that galaxies with a larger H$\alpha$ extent have larger width differences. This is probably because these galaxies have larger HI extents to match their H$\alpha$ extents, so the HI width traces further out into the halo, while we continue to measure the ORC width at the optical radius. The difference between the two measures of the rotation width also depends strongly on the slope of the rotation curve at the optical radius. \citet{C07} fit for this trend using 873 SFI++ galaxies for which both high quality HI widths and ORCs are available and find
\be
W_{21}/W_{\rm ORC}  =  \left\{ \begin{array}{ll}
	0.899 + 0.188 r_{\rm max}/r_{\rm opt} & \mbox{for rising ORCs} \\
	1.075 - 0.013 r_{\rm max}/r_{\rm opt} &	\mbox{for flat ORCs,} 
\end{array}
\right.
\label{ORCcorrection}
\ee
where $r_{\rm max}$ is the maximum extent of the ORC (\ie~ the H$\alpha$ extent of the galaxy) and the definition of a flat ORC is one in which the gradient of the rotation curve at the optical radius ($r_{\rm opt}$) is less than 0.5 \kms/arcsec; rising ORCs have gradients greater than this. For flat ORCs the correction increases the width by a fixed amount of $\sim$7\%, almost independent of $r_{\rm max}/r_{\rm opt}$. Rising ORCs have corrections ranging from increases of $\sim$15\% for galaxies with large H$\alpha$ extents, to decreases of $\sim$7\% for a small number of galaxies with very small H$\alpha$ extent. Once this correction is applied there are no systematic trends in the difference between HI and ORC widths, and the widths match within a mean scatter of $\sim 5$\%.

In the template sample there is a small trend of increasing H$\alpha$ extent with rotation velocity such that there are no small width galaxies with large H$\alpha$ extent, although large width galaxies span a range of H$\alpha$ extent. The ORC width correction therefore makes the TF slope for galaxies with ORC widths slightly shallower than the uncorrected version. It also slightly increases the zeropoint. The correction may also have an impact on the cluster peculiar velocities as derived in Section 6. Since the sky distribution of ORC widths vs. HI widths is quite uneven, some of the template clusters have virtually all galaxies with widths from HI measurements, while others are dominated by galaxies with ORC widths. A systematic difference in the template zeropoint for HI and ORC widths could therefore be absorbed in the cluster offsets (or peculiar velocities) calculated in Section 6 when the individual cluster samples are combined. We find no trends in the calculated offsets with the fraction of galaxies in a cluster which have ORC widths {\bf after} the correction is applied, providing reassurance that we are correcting for this effect properly.
 
\begin{figure} 
\epsscale{1}
\plotone{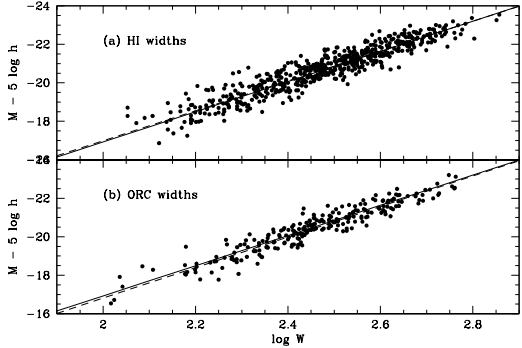}
\caption{TF diagrams for galaxies in the {\bf in+} sample. Panel (a) shows galaxies with widths derived from HI global profiles; Panel (b) shows galaxies with widths derived from optical rotation curves, the widths here have been corrected using Equation \ref{ORCcorrection}.  The solid line in both panels shows the bivariate fit to the total sample (see Section \ref{globalfit}), while the dashed lines show the bivariate fits to the sample separated by the source of the widths. The parameters of the fits are given in Table \ref{globalfits}. After the correction is applied to difference between the relations for the two subsamples is not significant.
\label{spec}}
\end{figure}

 Figure \ref{spec} shows the TF relation for galaxies in the template sample separated into galaxies with HI widths, and galaxies with widths from their ORC. The data have been corrected for all biases discussed in Section 3, and galaxies in each cluster sample have been shifted by the appropriate amount to account for the cluster peculiar velocity. As is also shown in Table \ref{globalfits} there is only a small difference in the TF fit to the two subsamples once the ORC widths have been corrected using Equation \ref{ORCcorrection}.  The difference is $\Delta M = -0.09 \pm 0.03 + (0.25 \pm 0.24)(\log W - 2.5)$. The zeropoints differ by $\sim 3 \sigma$, however we consider that this is not important due to the uneven sky distribution of the clusters. 

\section{Sample Selection and Bias Corrections} \label{bias}

\subsection{Sample Selection}
Clusters are chosen to be included in the TF template if they have more than $\sim$7 galaxies from the SFI++ data set in them. The Virgo cluster is not included in the sample due to the complicated nature of assigning galaxies to its various subgroups. Clusters too close to the Galactic plane are also not used, because of the uncertainty in Galactic extinction corrections. The final sample includes 22 of the 24 clusters used by G97b to derive the template for the SFI, and an additional 9 clusters for which we now have sufficient numbers of galaxies with data to include. The two missing G97b clusters are A2197 and A2666. Galaxies in Abell 2197 are absorbed into the periphery its partner cluster A2199 instead of being treated separately. A2666 is a small cluster, close in position but at slightly lower redshift than the template cluster A2634; we choose not to include 9 galaxies associated with A2666 because of the difficulty of disentangling them from the A2634 foreground or peripheral members. The assignment of galaxies to all clusters is discussed in Springob \etal ~(2007). We define two subsamples; 486 of the galaxies are considered {\it bona fide} cluster members, and are therefore assigned to the {\bf ``in"} sample. An additional 321 galaxies are considered peripheral cluster members and are added in what we call the {\bf ``in+"} sample. In deriving absolute magnitudes, we use the cluster redshifts for galaxies in the {\bf in} sample but the galaxy's own redshifts for those in the {\bf in+}. Galaxies in the {\bf in+} sample generally have recessional velocities very similar to that of the cluster so the choice between using the cluster or galaxy redshift makes little difference. An exception to this rule is the Cancer cluster \inp ~sample. Cancer is a complicated cluster with several distinct subgroups. Following G97b we consider only galaxies in the A clump of Cancer to be members of the \ins~ sample. Galaxies in the B, C, and D clumps are included in the \inp ~sample, and individual galaxy magnitudes are calculated using the recessional velocity of the relevant clump.

\begin{figure} 
\epsscale{1}
\plotone{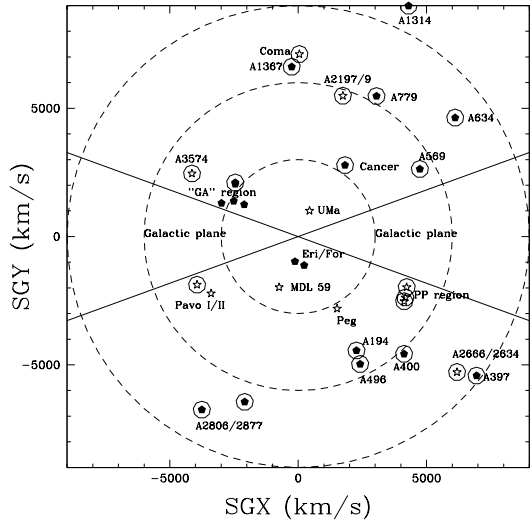}
\caption{The distribution of template clusters shown projected onto the supergalactic plane. Positions are shown in redshift space (CMB frame) with no correction for peculiar velocities. Clusters are shown as open stars if they lie above the SGP (\ie ~at +SGZ) and filled hexagons if they lie below it. Clusters used to define the ``reference frame'' in Section 5 are circled. The Pisces-Perseus (PP) region contains the nearby clusters NGC 383, NGC 507 and Abell 262. In the ``GA" region the five template clusters are Antlia, Hydra, NGC 3557, Centaurus 30 and ESO 508. The concentric dashed circles shown radii of 3000, 6000 and 9000 \kms. The Galactic plane ($\sim\pm20$\arcdeg) is indicated. \label{clustermap}}
\end{figure}

 Figure \ref{clustermap} shows the distribution of clusters used in the template in the supergalactic plane. The positions of the clusters are shown in redshift space (CMB frame) with no correction for peculiar velocity. Clusters are shown as open stars if they lie above the plane (\ie~ at +SGZ) and filled hexagons if they lie below it. Clusters used to define the ``reference frame'' in Section \ref{globalfit} are circled.

\subsection{Incompleteness Bias \label{cbias}}

\begin{figure*}
\epsscale{0.8}
\plotone{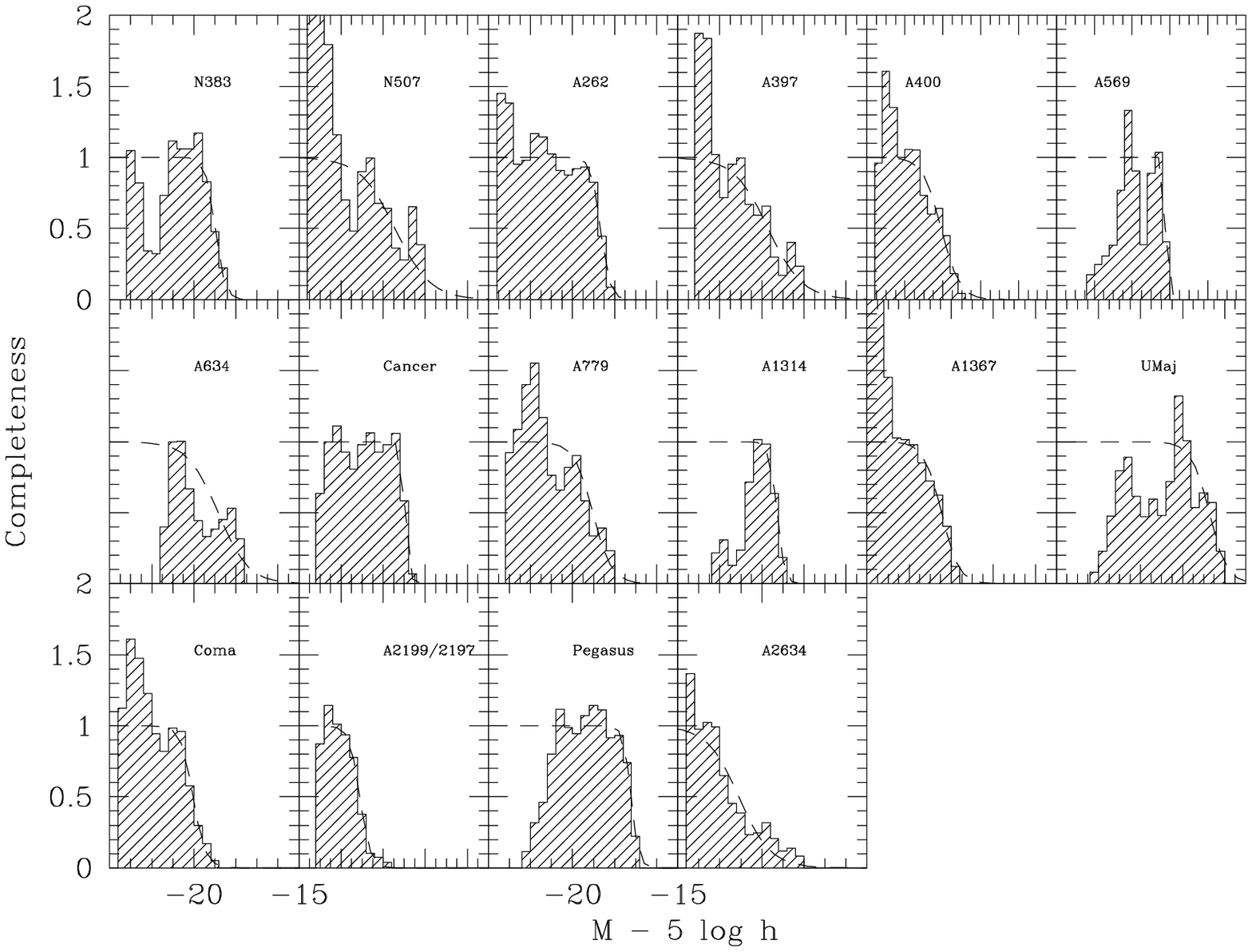}
\plotone{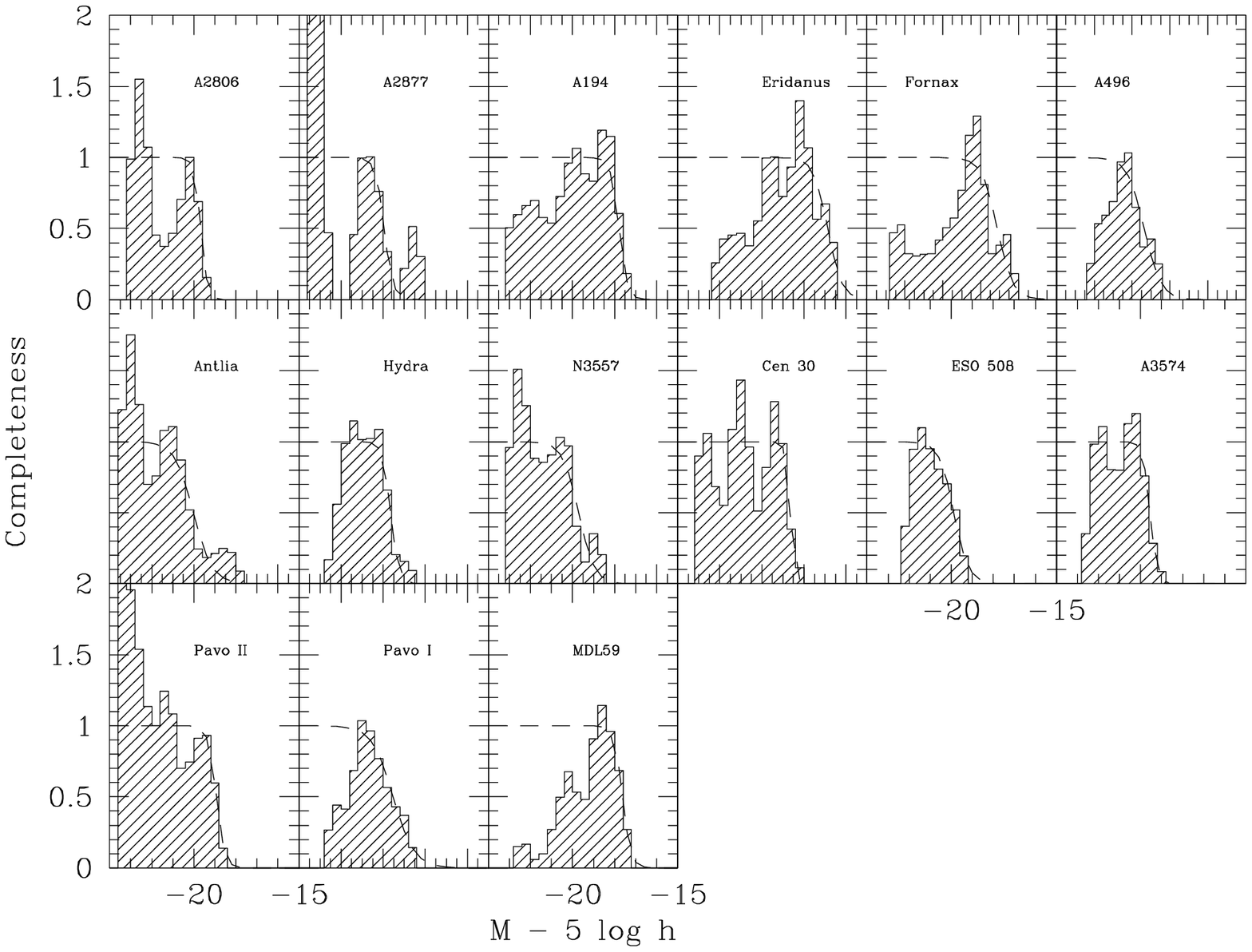}
\caption{Absolute magnitude completeness histograms for each of the {\bf in+} cluster sample. The dashed lines show the smooth fits to the histograms used in our Monte Carlo simulations. The template clusters are ordered by RA and separated into Northern Hemisphere (top) and Southern hemisphere (bottom) objects.
\label{clustercomplete}}
\end{figure*}

\begin{figure*} 
\epsscale{0.8}
\plotone{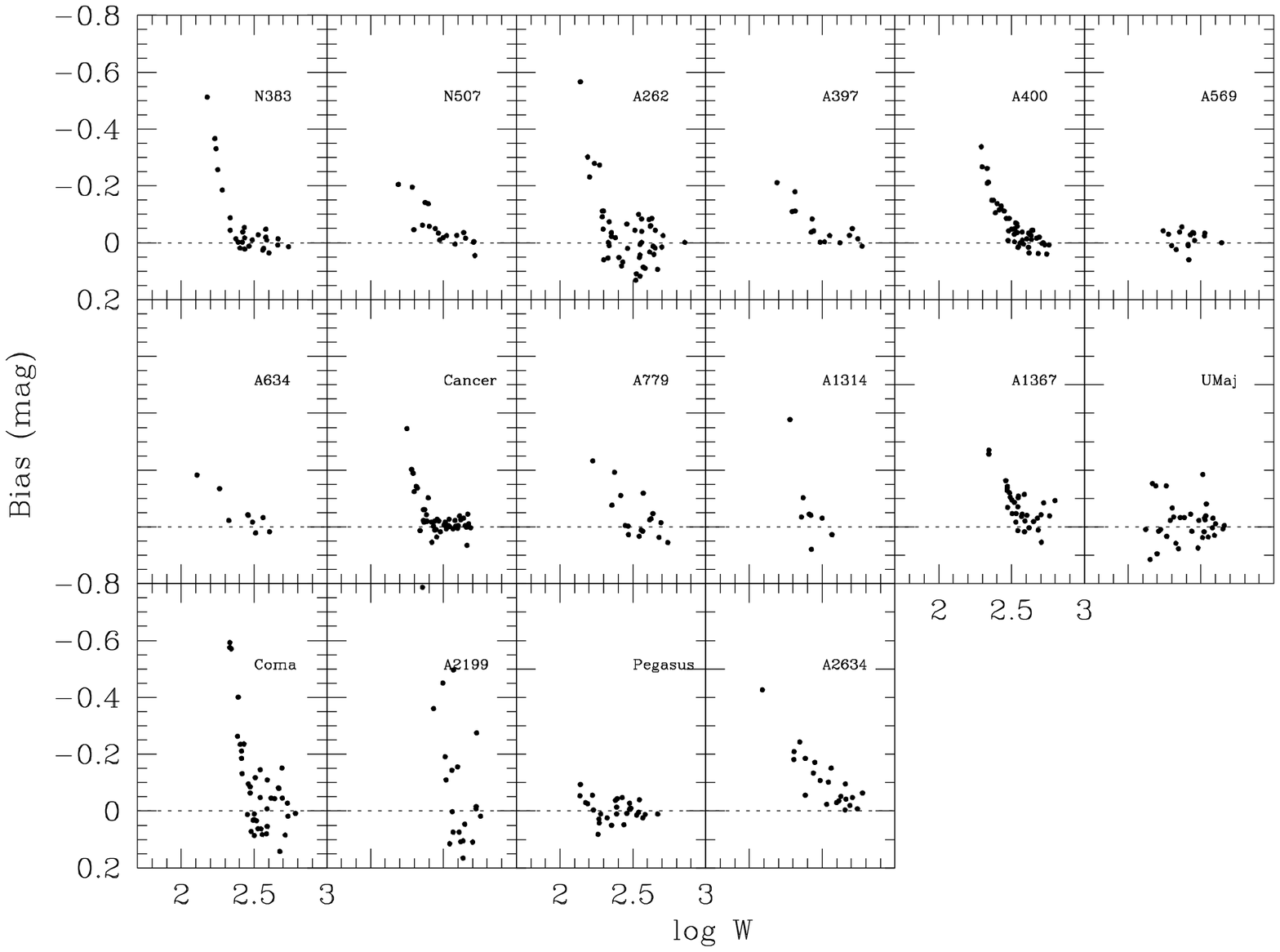}
\plotone{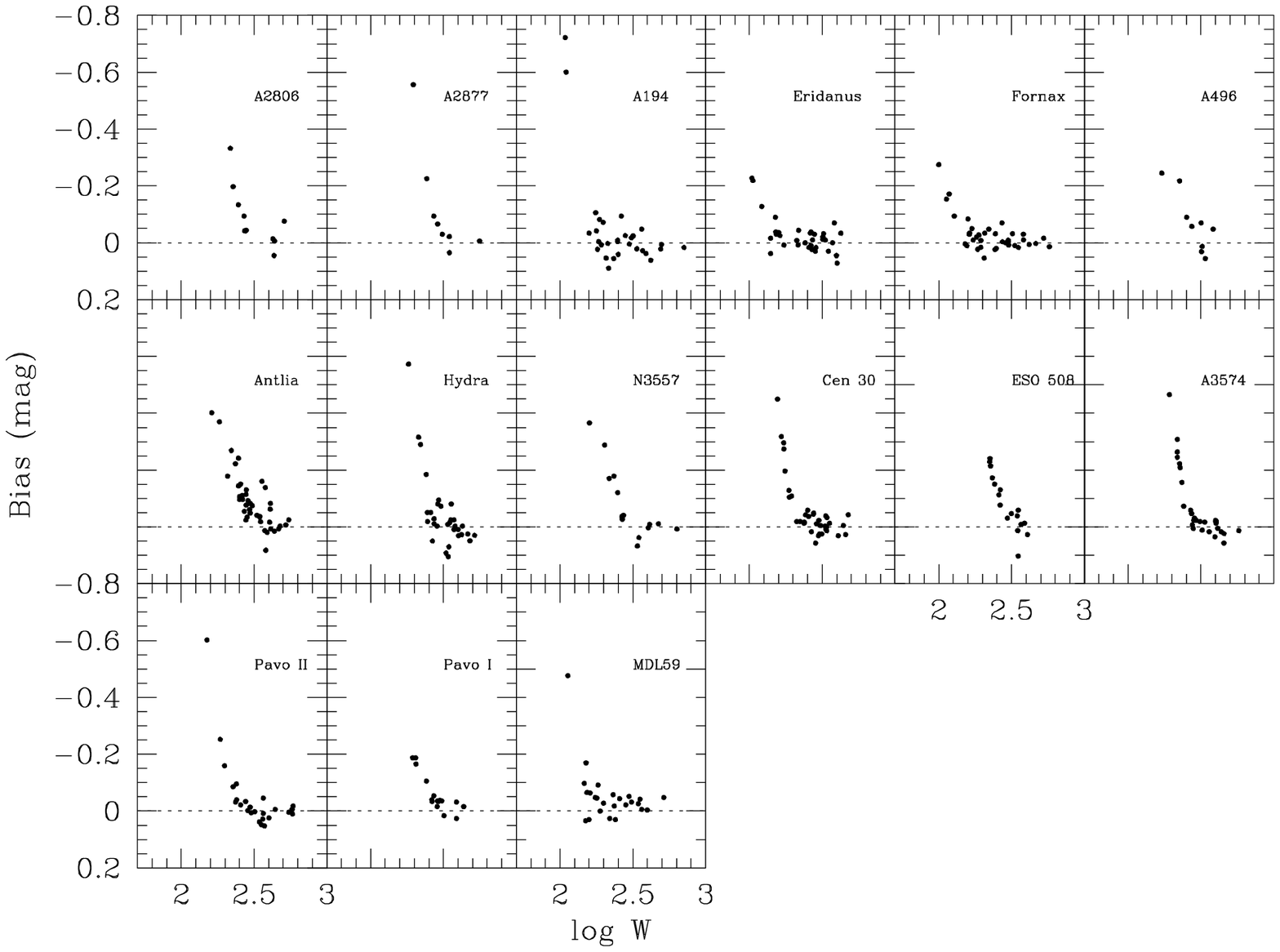}
\caption{Incompleteness bias correction for galaxies in each of the template clusters shown as a function of $\log W$.
\label{clustercors}}
\end{figure*}

As discussed in detail in G97b, the fact that all the galaxies in a cluster are at roughly the same distance, does not mean that a cluster sample is volume limited. Any implicit or explicit apparent magnitude limit in the sample means that in a given cluster we preferentially observe the brighter galaxies. Since the TF relation has an intrinsic scatter, this effect will tend to shallow the observed slope, by preferentially selecting galaxies which have brighter than expected magnitudes at the small width end of the relation. Similarly, it also brightens the zero point and decreases the observed scatter. In our treatment of this bias we follow G97b, and the reader is referred there for further explanation of the construction of completeness histograms for each of cluster samples and the Monte Carlo simulations used to derived the incompleteness bias from them. The completeness histograms for each cluster are shown in Figure \ref{clustercomplete} (note that in this Figure and elsewhere in the paper the clusters are listed in order of RA, and separated into Northern and Southern hemisphere objects). A function of the form 
\be
c(y) = \frac{1}{e^{(y-y_F)/\eta} + 1}
\label{cf}
\ee
is fit to these histograms; the parameters of these fits are shown in Table \ref{completeness} and it is these functions which are used as inputs for the bias calculations. The final incompleteness bias corrections for the galaxies in each cluster are shown in Figure \ref{clustercors}.

\begin{deluxetable*}{lcccccccc} 
\tablecolumns{9} 
\tablewidth{0pc} 
\tablecaption{Incompleteness Function Co-efficients (see Equation \ref{cf}) and cluster size bias (see Section \ref{sizebias}\label{completeness})}
\tablehead{ 
\colhead{Cluster Name} & \colhead{Redshift} & \colhead{N$_{\rm in}$}& \colhead{N$_{\rm in+}$}&\colhead{$y_F$} & \colhead{$\eta$}& \colhead{$\sigma_d/d$}& \colhead{{\bf in} $\beta_{\rm size}$} & \colhead{{\bf in+} $\beta_{\rm size}$}
}
\startdata 
N383 & 4865* &10 &27& -18.98 & 0.23 & 0.03 & -0.002 & -0.010\\
N507 & 4808* &14 &19& -19.43 & 0.83  & 0.03 & -0.002 & -0.010\\
A262 & 4665* &26&49& -18.66 & 0.22 & 0.03 & -0.003 & -0.010 \\
A397 & 9594* &14&15& -19.70 & 0.80 & 0.02 & -0.001 & -0.002 \\
A400 & 6934* &16&50& -20.08 & 0.48 &0.02 & -0.001 & -0.005 \\
A569 & 6011* &13&16& -18.22 & 0.05 & 0.02 & -0.001 & -0.006 \\
A634 & 7922* &9&-& -18.85 & 0.72 & 0.02 & -0.001 & - \\
Cancer & 4939* &17&49& -18.96 & 0.10 & 0.03 & -0.002 & -0.009 \\ 
A779 & 7211* &13&17&  -19.04 &0.45& 0.02 & -0.001 & -0.004 \\
A1314 & 9970* &8&-& -19.29 & 0.17 & 0.02 & -0.001 & - \\
A1367 & 6735* &32&33& -20.47 & 0.41 & 0.02 & -0.001 & -0.005 \\
Ursa Major & 1101 &28&34& -16.74 & 0.43 & 0.14 & -0.046 & -0.186 \\
Coma & 7185* &34&43& -20.03 & 0.11 & 0.02 & -0.001 & -0.004 \\
A2199/7 & 8996* &8&22& -21.11 & 0.25 & 0.02 & -0.001 & -0.003 \\
Pegasus & 3519 &19&30& -17.22 & 0.18 & 0.04 & -0.005 & -0.018 \\
A2634 & 8895* &18&22& -21.10 & 0.76 & 0.02 & -0.001 & -0.003 \\
\tableline
A2806 & 7867* &10&-& -19.52 & 0.06 & 0.02 & -0.001 & -\\
A2877 & 6974* &9&-& -19.95 & 0.21 & 0.02 & -0.001 & - \\
A194 & 5036* &23&31& -17.78 & 0.19 & 0.03 & -0.002 & -0.009\\
Eridanus & 1536  &21&34& -16.87 & 0.33 &0.10 & -0.024 & -0.095 \\
Fornax & 1321 &20&37& -17.12 & 0.41 & 0.11 & -0.032 & -0.129 \\
A496 & 9809* &9&-& -19.36 & 0.44 & 0.02 & -0.001 & - \\
Antlia & 3120 &17&41& -20.01 & 0.45 & 0.05 & -0.006 & -0.023 \\
Hydra & 4075* &21&31& -19.64 & 0.19 & 0.04 & -0.003 & -0.014 \\
NGC 3557 & 3120 &7&14& -19.71 & 0.37 &0.05 & -0.005 & -0.020 \\
Cen 30 & 3322 &23&36& -18.65 & 0.11 & 0.05 & -0.005 & -0.020 \\
ESO 508 & 3196 &9&17& -19.87 & 0.35 & 0.05 & -0.006 & -0.022 \\
A3574 & 4817*  &13&29& -19.56 & 0.17 & 0.03 & -0.002 & -0.010 \\
Pavo II & 4444* &8&25& -18.93 & 0.19 & 0.03 & -0.003 & -0.011 \\
Pavo I & 4055 &5&16& -19.52 & 0.50 & 0.04 & -0.003 & -0.014 \\
MDL59 & 2317 &12&25& -17.62 & 0.20 & 0.07 & -0.010 & -0.042
\enddata
\tablenotetext{*}{Cluster used to define the rest frame for peculiar velocities in Section 5.1.}
\end{deluxetable*}

 The incompleteness correction derived here relies on a priori knowledge of the intrinsic TF relation and scatter, and as such proceeds in an iterative manner. The magnitude of the assumed scatter has by far the biggest impact (when the scatter is larger the bias corrections are larger and vice versa), so the derived corrections change very little once the intrinsic scatter has been determined. The impact that the various assumptions have on the derived corrections is illustrated in Figures 7, 8 and 9 of G97b. As well as requiring an input TF relation, the derivation of this bias correction must assume knowledge of the spiral luminosity function (LF).  As shown in \cite{BST88}, the spiral LF has an almost Gaussian shape. Following G97b we use a Schechter function with a steep faint end slope, $\alpha = -0.5$, and $M_\star = -21.6$.  This assumption impacts the derived bias corrections because it determines the level of incompleteness in the sample.  If the LF slope is in actual fact shallower than this the bias will be underestimated (and vice versa).
  
 We perform several iterations of the bias correction, until the final TF relation differs by less than 1$\sigma$ from the input values. The final input TF relation is $M - 5 \log h = -20.85 - 7.81(\log W - 2.5)$ with intrinsic scatter of $\epsilon = 0.36 - 0.34 (\log W - 2.5)$. The final result is $M - 5 \log h = -20.85 - 7.85(\log W - 2.5)$ and $\epsilon = 0.35 - 0.37 (\log W - 2.5)$ as discussed in Sections 5 and 6.

\subsection{Cluster size bias} \label{sizebias}
In deriving the template we are explicitly assuming that all galaxies in a given cluster are at the same distance. This assumption varies in its accuracy based on the actual line of sight depth of the cluster relative to its distance The finite size of the cluster introduces two separate biases into the derived TF relation:
\begin{trivlist}
\item{\it 1. The mean distance to a cluster.}
Fitting a TF relation under the assumption that all galaxies in a cluster are at the same distance, is equivalent to fitting the mean of the $\log$ of the distance to the cluster (because we work with distance modulus rather than distance). The $\log$ of the average is not the same as the average of the $\log$, so this assumption introduces a small bias into the zeropoint of 
\be
\beta_{\rm size} = M_{\rm true} - M_{\rm calc} =  - \frac{5 \sigma_d^2}{2 d^2},
\ee
where $\sigma_d$ represents the line of sight depth of the cluster at a distance $d$. Table \ref{completeness} lists the size of this bias for the clusters in our sample, using the assumption that they all have a line of sight extent of 1 Abell radius (\ra, 1.5 Mpc/h or 150 \kms). This is a reasonable approximation for most of the \ins~ sample. For the {\bf in+} sample, we use twice this extent, increasing the small bias by a factor of four. 
\item{\it 2. Cluster size -- sample incompleteness.}
   If galaxies are close in magnitude to the completeness limit of the sample they will preferentially be seen only in the foreground parts of the cluster (\ie~ the closest parts) and therefore their magnitudes will be biased towards brighter values. This bias is illustrated in Figure 14 of G97b, and is estimated assuming a Gaussian extent for the cluster, and using the completeness functions derived in Section \ref{cbias}. This bias is very small (less than 0.1 mag) for most galaxies in our sample.
\end{trivlist}

\subsection{Edge of Catalog Bias and Homogeneous Malmquist Bias}
Following G97b we make no correction for either the edge of catalog bias or the Malmquist bias, which they find to be negligible in the SCI sample on which this sample is based. 

\subsection{Morphological Dependence of the TF Relation}
\begin{figure} 
\epsscale{1}
\plotone{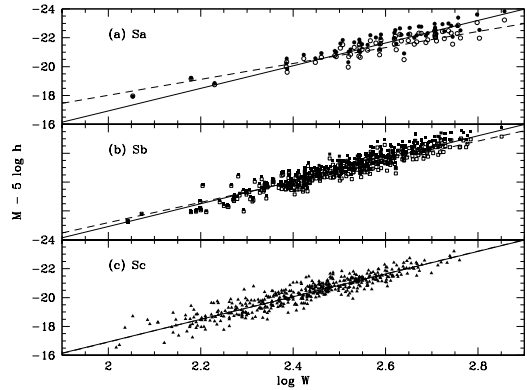}
\caption{TF diagrams for galaxies in the {\bf in+} sample. (a) galaxies with types earlier than Sb; (b) galaxies of type Sb; (c) galaxies with types later than Sb. The solid line in all three panels shows the bivariate fit to the {\bf in+} sample, while the dashed lines show the bivariate fits to the sample separated by morphological type. The slopes vary from -5.52 for types earlier than Sb to $-7.87$ for types later than Sb. The parameters of the fits are given in Table \ref{globalfits}. Open symbols show the magnitudes corrected for all biases except morphological type, while the filled symbols show the magnitudes after the morphological type correction has been applied.
\label{morph}}
\end{figure}

Figure \ref{morph} shows TF diagrams of galaxies in the {\bf in+} sample separated into three groups by their morphological type. These data have been corrected for all biases discussed above and galaxies in each cluster sample have been shifted by an amount $\Delta M$ to account for the cluster's peculiar velocity as discussed in Section \ref{globalfit}. We find that earlier types have a shallower TF slope than later types as also noted in G97b. The relation for Sa galaxies is shallower than that for Sc galaxies by an amount $2.35\pm0.43$ mag/dex, while the relation for Sb galaxies is shallower than for Scs by an amount $0.80\pm0.22$ mag/dex. (see Table \ref{globalfits} for the full fit parameters). We need to be careful in interpreting these results, because the completeness of galaxies of the different types is different in such a way that the incompleteness bias may be larger for the earlier type galaxies (as is obvious in Figure \ref{morph}). If true, this would artificially shallow the slope of their relation, perhaps accounting for the observed difference. In fact in the SFI sample an effort {\bf was} made to be more complete for later type galaxies (this constraint was relaxed for the additional galaxies added to the SFI++). Nature also conspires to make the completeness of later type spirals better in a TF sample since later type galaxies are more likely to be strong HI or H$\alpha$ emitters making the measurement of rotation widths easier. However,  the luminosity function of early type spirals is such that there are intrinsically fewer low luminosity objects \citep{BST88} which could also explain the lack of small width Sa galaxies in the sample without the need for greater incompleteness for earlier type spirals.

 The observed difference in slope between morphological types (which is significant after a mean incompleteness bias has been subtracted) can therefore either be interpreted as a real physical difference between different types of spiral galaxies, or as an indication that the incompleteness bias is an underestimate for the earlier type spirals. Without a more extensive study of the luminosity functions for different spiral types there is no easy way to disentangle the two effects. The fact that the biggest difference is seen at the high width end of the relation where the bias corrections are smallest (see Figure 5) however argues for a physical explanation. In either case the difference should be corrected for in the template. Given the small number of Sas in the sample (61) and the uncertainty in deriving the slope of the TF relation, we will correct for this effect for all 342 Sa/Sb spirals together (although for reference the relations are given for all 3 groups separately as well). We first correct for a plain additive offset, found by fitting all samples with the Sc slope. This correction is identical to that found by G97b (see below). After this correction we find the TF relation for the combined Sa/Sb sample (bivariate fit) to be $M - 5 \log h = -20.85(2) -6.92(15)(\log W - 2.5)$, shallower than the Sc relation by an amount $0.95\pm0.21$ mag/dex. We therefore apply the following corrections to the magnitudes:
\begin{itemize}
\item S0/Sa/Sab: $-0.32 -0.9 (\log W - 2.5)$  mag
\item Sb: $-0.10 -0.9 (\log W - 2.5)$ mag
\item Sbc/Sc/Scd: no correction.
\end{itemize}

 If the difference in slope for the different morphological types of spiral galaxies cannot be explained by differing amounts of incompleteness then it is telling us something about the physical differences between early and late type spirals. It is already well known that early type spirals are dimmer than late types at a given rotational speed, what the dependence of the TF slope on morphological type shows is that this difference gets more pronounced the more massive the galaxies are. The slope of the TF relation is fairly well reproduced in numerical simulations of disk galaxy formation \citep{M00,NS00}. \citet{M00} suggest that the slope could be shallower than the fiducial $L \propto v^3$ (observed here for Scs) if galaxies at the high-width end of the relation have systematically less concentrated halos. \citet{NS00} show that under the assumption of constant disk mass-to-light ratio, $L \propto v_{\rm rot}^\alpha$ implies that the fraction of mass in the disk, $f_{\rm disk} \propto (v_{\rm rot}/v_{\Delta})^\alpha$ (where $v_\Delta$ is the circular velocity of the dark matter halo at the virial radius).  The observed difference in TF slope for different morphological types indicates therefore that early type spirals must have a smaller $f_{\rm disk}$ at a given rotational velocity, presumably because they have mass in a bulge component (we find $L \propto v^{3.1}$ for Sbc/Sc/Sd, $L \propto v^{2.8}$ for Sbs and $L \propto v^{2.2}$ for Sab/Sa/S0s). Furthermore \citet{NS00} shows that halo concentration must increase with $f_{\rm disk}$ to reproduce the observed TF relation; the steeper the slope of the relation the more quickly the concentration increases with $f_{\rm disk}$. Putting this together it suggests that as well as being intrinsically dimmer earlier type spirals may have systematically less concentrated halos than later types at a given rotational velocity.

\subsection{Summary of all Bias Corrections}
 Here we summarize for clarity all bias corrections which are applied to the data in advance of calculating the final TF relations and scatter. These are corrections applied {\it after} the magnitudes have been corrected for Galactic and internal extinction and the inclinations have been corrected for seeing (see Section 2.1), and after the HI widths have been corrected for instrumental broadening and all widths corrected for inclination and cosmological broadening (see Section 2.2).
\begin{itemize}
\item[{\it 1.}] {\it ORC widths corrected to HI scale (Section 2.2).} HI widths are observed to be on average larger than ORC widths for the same galaxy. This correction is derived by \citet{C06} using 873 SFI++ galaxies which have both HI widths and ORCs.
\item[{\it 2.}] {\it Morphological correction (Section 3.5).} Earlier type spirals are observed to be on average dimmer than later type spirals at the same width. We also notice a significant shallowing of the TF slope for earlier type spirals. The slope correction applied is the same for all spirals earlier than Sc, while the fixed offset varies from -0.10 mag for Sb galaxies to -0.32 mag for spirals earlier than Sb.
\item[{\it 3.}] {\it Incompleteness bias correction (Section 3.2).} In any given cluster we preferentially observe galaxies which are scattered above the TF slope (because they are brighter). This has a tendency to shallow the TF slope, brighten the zero point and reduce the scatter. The correction we apply is derived from Monte Carlo simulations requiring an input TF relation, TF scatter, and assumptions about the spiral LF. The calculation is done several times until convergence is reached between the input and final TF.  
\item[{\it 4.}] {\it Mean distances to cluster (Section 3.3.1).} This small fixed bias for all galaxies in a given cluster arises from the fact that the $\log$ of the average of the distance is not equal to the average of the $\log$ of the distance (which is what we implicitly fit for when using magnitudes). This bias is easy to calculate once a line of sight depth for each cluster has been estimated. 
\item[{\it 5.}] {\it Cluster size -- incompleteness (Section 3.3.2).} In a given cluster we preferentially observe dimmer galaxies only in the closest parts of the cluster - creating a small bias in the observed magnitudes which is easily corrected for using the same assumed completeness as in Section 3.2 and assumptions about the size of the cluster.
\item[{\it 6.}] {\it Edge of catalog bias and Malmquist bias.} We argue that both of these biases are negligible in the sample and therefore make no correction for them.
\item[{\it 7.}] {\it Cluster peculiar velocity (Section 5).} To combine all the individual cluster samples into a global  TF relation, a peculiar velocity must be estimated for each cluster. This is done by minimizing the scatter in the TF relation, and the absolute value of the zeropoint is set by assuming that the net motion of a subset of the most distant clusters is equal to zero. 
\end{itemize}

\section{Individual Cluster TF Relations and the Impact of Environment}\label{indivTF}

Here we fit the TF relation, first to each template cluster sample individually. We fit ``direct'', ``inverse'', ``bivariate'' forms of the linear TF relation. The difference between the three linear fits is in which variable is considered to be independent and the treatment of the errors. For more details of the fitting procedure see G97b.

\begin{figure*} 
\epsscale{0.8}
\plotone{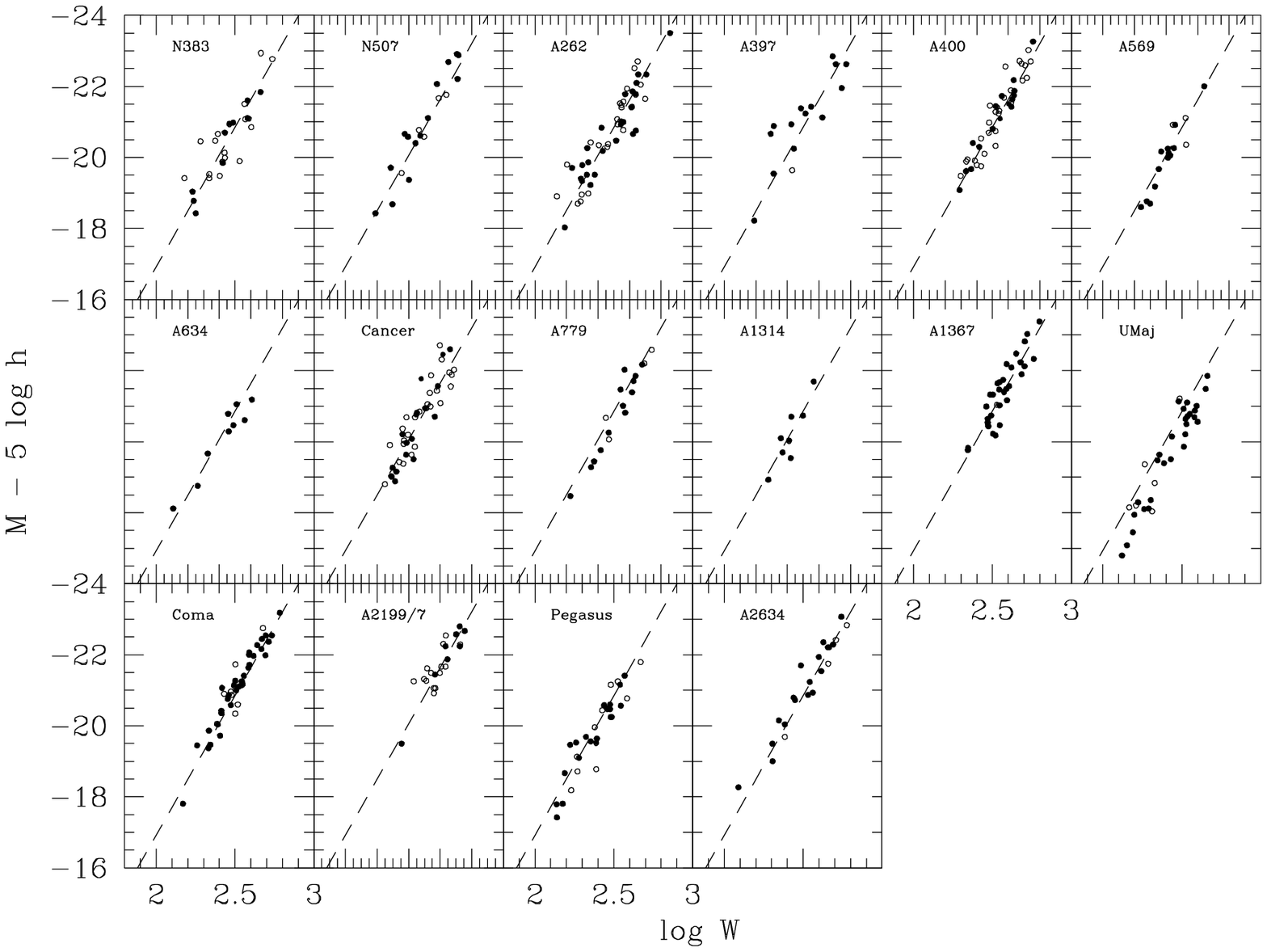}
\plotone{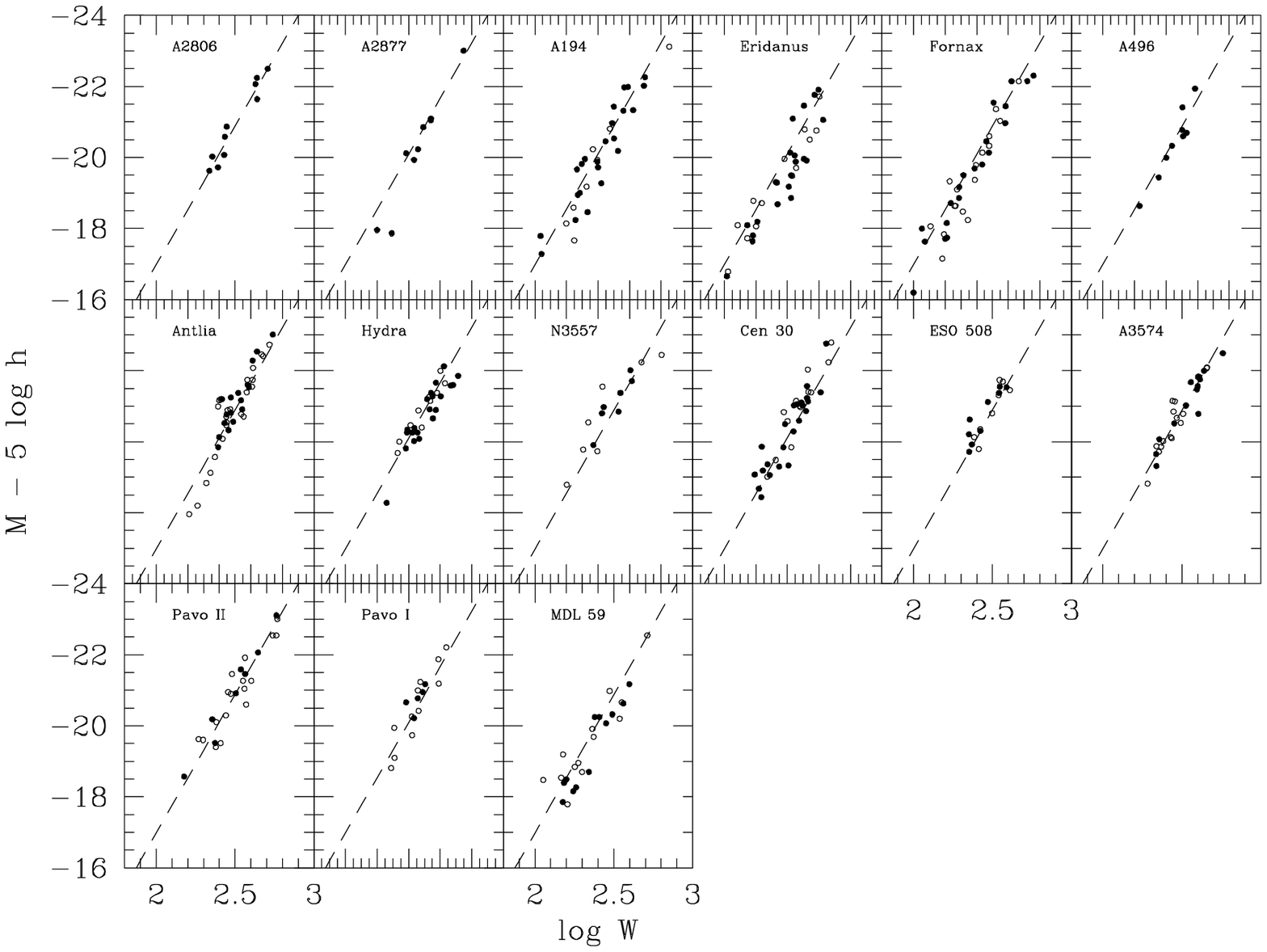}
\caption{TF relations for individual clusters (not corrected for cluster peculiar velocities). Overlaid is the bivariate fit to the {\bf in} sample from Section \ref{globalfit} of $M - 5 \log h = -20.849 - 7.845 (\log W -2.5)$. 
Solid circles are {\bf in} sample galaxies, open circles {\bf in+}. These data have been corrected for the incompleteness bias and other biases discussed in Section 3.
\label{clusterTFs}}
\end{figure*}

The individual template cluster TF relations are shown in Figure \ref{clusterTFs} . Galaxies in the {\bf in} and {\bf in+} samples of each cluster are plotted as solid and open circles respectively. The same line showing the bivariate fit to the combined sample is plotted over each of the cluster sub-plots. The parameters of the fits to individual clusters samples are given in Table \ref{clusterfits}. 

\begin{deluxetable*}{lccccccccccc} 
\tablecolumns{12} 
\tablewidth{0pc} 
\tabletypesize{\footnotesize}
\tablecaption{Tully-Fisher relation fits for individual cluster samples \label{clusterfits} }
\tablehead{ 
\colhead{Sample} & \colhead{$N$}   & \colhead{$a_{dir}$}    & \colhead{$b_{dir}$} & 
\colhead{$a_{inv}$} & \colhead{ $b_{inv}$}   & \colhead{$a_{bi}$}    & \colhead{$b_{bi}$} & 
\colhead{$\epsilon_a$}    & \colhead{$\epsilon_b$}   & \colhead{$\sigma_{bi}$}    & \colhead{$\sigma_{abs}$}}
\startdata 
N383      \inp &   27& -20.827&  -7.623& -20.858&  -9.988& -20.844&  -8.324&   0.080&   0.612&   0.538&   0.440\\
N383      \ins  &   10& -20.817&  -8.261& -20.782&  -9.035& -20.807&  -8.596&   0.135&   1.030&   0.318&   0.295\\
N507      \inp &   19& -20.995&  -8.743& -21.103&  -9.102& -21.015&  -8.750&   0.091&   0.596&   0.393&   0.335\\
N507      \ins  &   14& -21.037&  -8.761& -21.167&  -9.259& -21.040&  -8.799&   0.104&   0.631&   0.431&   0.367\\
A262      \inp &   49& -20.849&  -7.423& -20.809&  -8.885& -20.821&  -7.687&   0.060&   0.382&   0.472&   0.383\\
A262      \ins  &   26& -20.791&  -7.367& -20.693&  -8.504& -20.739&  -7.733&   0.084&   0.496&   0.467&   0.371\\
A397      \inp &   15& -21.074&  -6.571& -20.720&  -8.972& -20.991&  -6.983&   0.121&   0.690&   0.585&   0.489\\
A397      \ins  &   14& -21.101&  -6.315& -21.023&  -7.397& -21.037&  -6.709&   0.125&   0.688&   0.574&   0.482\\
A400      \inp &   50& -20.943&  -8.304& -20.862&  -9.066& -20.923&  -8.336&   0.062&   0.478&   0.334&   0.271\\
A400      \ins  &   16& -20.882&  -8.384& -20.932&  -9.468& -20.850&  -8.586&   0.111&   0.811&   0.292&   0.258\\
A569      \inp &   16& -20.840&  -8.915& -20.755&  -9.476& -20.839&  -8.984&   0.143&   1.082&   0.320&   0.238\\
A569      \ins  &   13& -20.853&  -9.144& -20.866&  -9.077& -20.863&  -9.194&   0.177&   1.178&   0.246&   0.206\\
A634      \ins  &    9& -20.615&  -6.506& -20.763&  -7.861& -20.650&  -6.771&   0.156&   1.389&   0.261&   0.222\\
Cancer    \inp &   49& -20.901&  -8.273& -20.914&  -9.738& -20.899&  -8.538&   0.060&   0.480&   0.462&   0.371\\
Cancer    \ins  &   17& -20.955&  -9.897& -21.039&  -9.776& -21.028& -10.468&   0.115&   0.898&   0.461&   0.345\\
A779      \inp &   17& -20.674&  -8.696& -20.567&  -9.801& -20.655&  -8.779&   0.110&   0.793&   0.314&   0.236\\
A779      \ins  &   13& -20.639&  -9.223& -20.586&  -9.212& -20.622&  -9.174&   0.123&   0.943&   0.314&   0.212\\
A1314     \ins  &    8& -20.917&  -9.605& -20.884&  -9.428& -20.951& -10.285&   0.234&   2.273&   0.334&   0.251\\
A1367     \inp &   33& -20.882&  -8.497& -20.598& -11.876& -20.813&  -9.177&   0.099&   0.775&   0.401&   0.329\\
A1367     \ins  &   32& -20.886&  -8.483& -20.558& -12.066& -20.815&  -9.166&   0.102&   0.780&   0.405&   0.336\\
Ursa Major \inp &   34& -20.329&  -8.508& -20.536&  -8.689& -20.343&  -8.627&   0.078&   0.484&   0.461&   0.376\\
Ursa Major \ins  &   28& -20.271&  -8.943& -20.484&  -8.980& -20.286&  -8.941&   0.081&   0.541&   0.406&   0.332\\
Coma      \inp &   43& -21.015&  -8.091& -20.998&  -8.393& -21.026&  -8.148&   0.069&   0.492&   0.307&   0.238\\
Coma      \ins  &   34& -20.986&  -7.988& -20.935&  -7.827& -20.994&  -8.011&   0.080&   0.539&   0.259&   0.200\\
A2199/7   \inp &   22& -21.077&  -7.178& -20.917&  -8.769& -20.948&  -8.215&   0.153&   1.025&   0.356&   0.310\\
A2199/7   \ins  &    8& -20.809&  -8.666& -20.802&  -8.819& -20.739&  -9.071&   0.287&   1.462&   0.243&   0.218\\
Pegasus   \inp &   30& -20.659&  -7.497& -21.109&  -9.826& -20.687&  -7.723&   0.099&   0.620&   0.405&   0.333\\
Pegasus   \ins  &   19& -20.683&  -7.774& -20.916&  -9.426& -20.706&  -7.978&   0.133&   0.769&   0.374&   0.297\\
A2634     \inp &   22& -20.993&  -7.850& -20.906&  -8.564& -20.936&  -8.129&   0.095&   0.562&   0.339&   0.267\\
A2634     \ins  &   18& -20.997&  -7.977& -20.932&  -8.146& -20.935&  -8.197&   0.102&   0.625&   0.323&   0.241\\
\tableline
A2806     \ins  &   10& -20.901&  -8.359& -20.781&  -8.883& -20.829&  -8.761&   0.159&   1.183&   0.242&   0.208\\
A2877     \ins  &    9& -20.627& -10.162& -20.776&  -9.547& -20.640& -10.405&   0.169&   1.184&   0.319&   0.209\\
A194      \inp &   31& -20.727&  -8.132& -20.954&  -9.670& -20.701&  -8.366&   0.085&   0.469&   0.491&   0.384\\
A194      \ins  &   23& -20.748&  -8.203& -21.053& -11.730& -20.729&  -8.663&   0.104&   0.718&   0.508&   0.413\\
Eridanus  \inp &   34& -20.466&  -7.942& -20.197&  -7.772& -20.411&  -7.935&   0.095&   0.519&   0.549&   0.469\\
Eridanus  \ins  &   21& -20.481&  -8.535& -19.847&  -5.485& -20.355&  -8.218&   0.124&   0.689&   0.588&   0.495\\
Fornax    \inp &   37& -20.572&  -8.357& -20.704&  -8.750& -20.608&  -8.765&   0.090&   0.455&   0.482&   0.380\\
Fornax    \ins  &   20& -20.538&  -8.070& -20.681&  -8.768& -20.565&  -8.575&   0.113&   0.544&   0.501&   0.399\\
A496      \ins  &    9& -20.870&  -9.042& -20.701& -12.329& -20.848&  -9.512&   0.150&   1.961&   0.308&   0.250\\
Antlia    \inp &   41& -20.969&  -8.694& -20.887& -10.899& -20.938&  -8.845&   0.071&   0.573&   0.415&   0.308\\
Antlia    \ins  &   17& -21.052&  -8.400& -21.181&  -9.018& -21.005&  -8.809&   0.121&   1.149&   0.321&   0.250\\
NGC 3557  \inp &   14& -21.072&  -6.269& -21.094&  -6.330& -21.084&  -6.293&   0.113&   0.681&   0.414&   0.341\\
NGC 3557  \ins  &    7& -21.070&  -6.850& -21.038&  -8.910& -21.073&  -6.983&   0.161&   1.810&   0.288&   0.268\\
Hydra     \inp &   31& -20.769&  -7.058& -20.757&  -9.699& -20.782&  -7.509&   0.076&   0.765&   0.323&   0.278\\
Hydra     \ins  &   21& -20.686&  -6.797& -20.751& -10.253& -20.704&  -7.338&   0.094&   0.918&   0.341&   0.302\\
Cen 30    \inp &   36& -21.100&  -8.375& -20.909& -10.008& -21.119&  -8.873&   0.077&   0.640&   0.414&   0.334\\
Cen 30    \ins  &   23& -21.009&  -8.025& -21.073& -11.684& -21.040&  -8.731&   0.108&   0.871&   0.439&   0.350\\
ESO 508   \inp &   17& -21.008&  -7.368& -21.145&  -8.054& -21.022&  -7.544&   0.107&   1.167&   0.298&   0.241\\
ESO 508   \ins  &    9& -21.085&  -6.967& -21.351&  -7.284& -21.095&  -6.996&   0.156&   1.502&   0.255&   0.213\\
A3574     \inp &   29& -20.878&  -7.492& -20.720&  -8.491& -20.873&  -7.466&   0.080&   0.625&   0.308&   0.220\\
A3574     \ins  &   13& -20.840&  -7.272& -19.812& -15.571& -20.827&  -7.328&   0.127&   0.907&   0.304&   0.197\\
Pavo II   \inp &   25& -20.925&  -7.776& -20.969&  -7.643& -20.929&  -7.861&   0.088&   0.526&   0.376&   0.302\\
Pavo II   \ins  &    8& -20.986&  -7.982& -21.123&  -7.492& -20.993&  -8.006&   0.157&   0.879&   0.204&   0.152\\
Pavo I    \inp &   16& -21.014&  -8.553& -20.653&  -7.942& -20.999&  -8.409&   0.115&   1.163&   0.355&   0.303\\
Pavo I    \ins  &    5& -21.006&  -6.100& -21.420& -14.862& -21.009&  -6.443&   0.289&   4.902&   0.220&   0.160\\
MDL59     \inp &   25& -20.549&  -7.222& -20.745&  -9.193& -20.568&  -7.623&   0.108&   0.550&   0.492&   0.404\\
MDL59     \ins  &   12& -20.401&  -7.437& -20.424&  -8.019& -20.411&  -7.571&   0.166&   0.949&   0.361&   0.286
\enddata
\end{deluxetable*}

\subsection{Effect of Environment on the TF relation}
It is well known that environment has a strong impact on galaxy evolution. Gas rich galaxies which fall into clusters are thought to be stripped of their gas and dynamically disturbed (be it by ram pressure stripping, galaxy-galaxy interactions or the tidal effects of the cluster potential), therefore it is possible that the TF relation of galaxies in clusters may differ from that in the field. It should be noted here that in correcting for the observed morphological type dependence of the TF relation we may already have removed much of the impact of environment. The observed morphology-density relations shows that early type spirals (which we observe to have a shallower TF slope than late types) are more likely to be found in more over-dense regions.

 We perform two simple tests to search for the effect of environment in our sample after correcting for morphological type. The first is to compare the slope, zeropoint and scatter of individual cluster samples as a function of global cluster properties. The X-ray selected cluster sample HIFLUGCS \citep{RB02} contains 12 of the 31 clusters in the template sample. For these clusters no significant trends are observed of the difference between the cluster and global slope or zeropoint with X-ray luminosity. There appears to be a slight tendency for clusters with higher L$_X$ to have smaller than average scatter in their TF relation, however this may be a result of a Malmquist bias like effect. The high L$_X$ clusters are preferentially at large distances and therefore have a relatively small number of galaxies with TF measurements.  It is observed here the TF scatter decreases as the number of galaxies in a cluster sample decreases, indicating non-Gaussian behavior in the scatter. As a second test for environmental effects we plot the residuals from the global TF relation as a function of the galaxies' projected cluster distance (Figure 7). Again no significant trends are observed. Both of these tests suggests that there is no reason to doubt that a global TF relation applies to all spirals galaxies, regardless of environment, once a correction for morphological type has been applied. 

\begin{figure} 
\epsscale{1}
\plotone{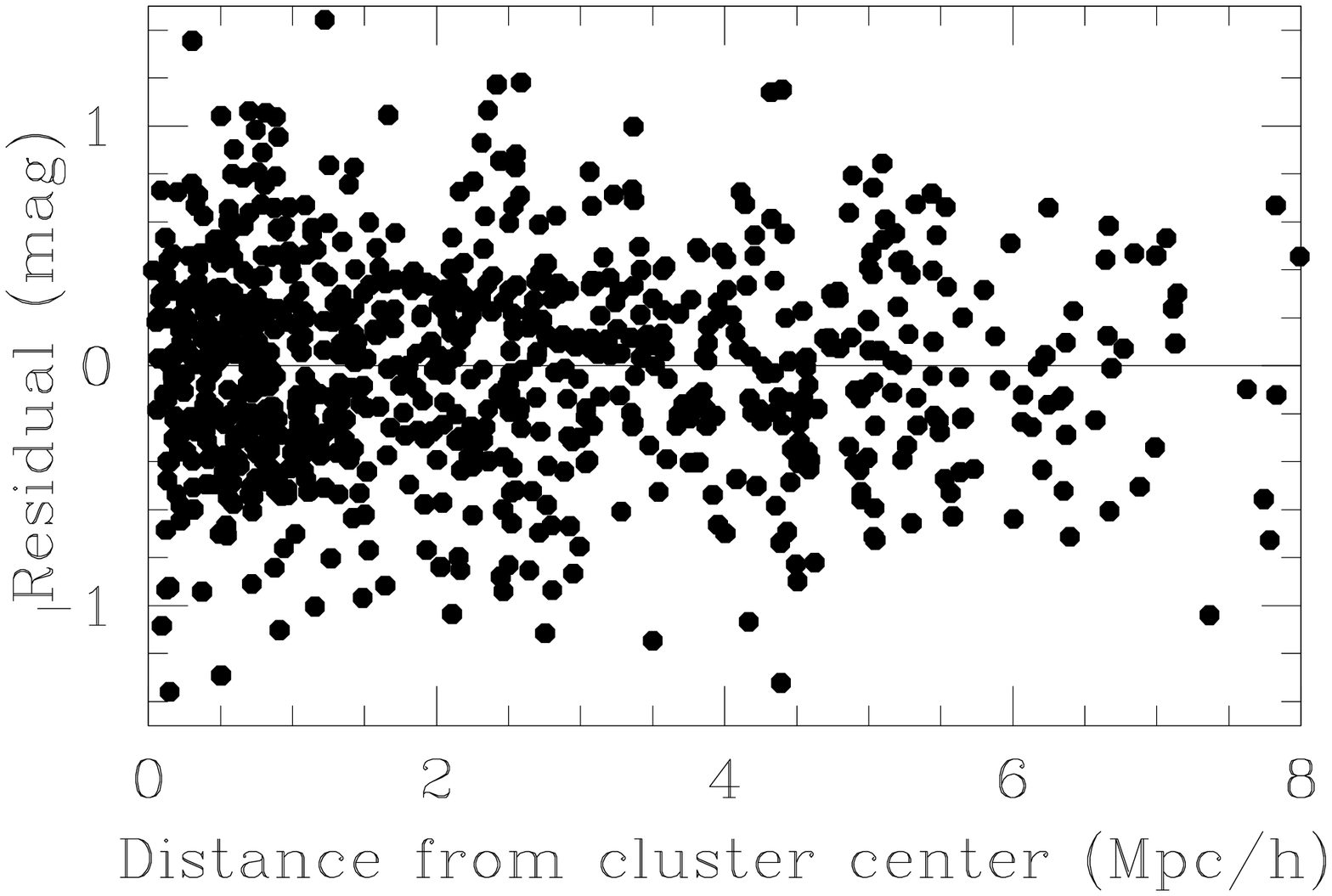}
\caption{Residual from the global TF relation (bivariate fit to the \inp~ sample) for all galaxies in the \inp ~ sample as a function of their projected distance from their cluster center. \label{residdist}}
\end{figure}

\section{A Global TF Template Relation and the Cluster Peculiar Velocity Dispersion} \label{globalfit}
 
 Here we combine the data from all 31 clusters into a global TF template. This is done by shifting the points for each individual cluster by an amount
 \be 
 \Delta M = 5 \log \left( \frac{v_{\rm CMB} - v_{\rm pec}}{v_{\rm CMB}} \right),
 \ee
  taking into account the motion of each cluster in the CMB frame (\ie~ its peculiar velocity). These offsets are found by assuming that all points fall on a linear TF relation with a global slope, and minimizing $\chi^2$ with respect the offsets. The global slope is assumed to be the bivariate fit to the combined sample and is found by iterating until the fit converges. Additionally an inertial reference frame is defined by requiring that the net motion of a subset of the most distant clusters (all those with $cz > 4000$ \kms, excluding the Pavo Group which has only a small number of galaxies) is zero, \ie,
\be
\sum_k N_k \Delta M_k / \sum_k N_k = 0
\ee
This sum is weighted by the number of galaxies in the cluster TF sample. The total number of galaxies in the reference clusters is 520 in the \inp ~sample, or 325 in \ins.  

\subsection{Cluster Peculiar Velocity Dispersion}
The motions of clusters probe the peculiar velocity field on a scale which is well within the linear regime where theory and observation can be linked relatively easily. The large number of galaxies in each cluster also allow the peculiar velocities of clusters to be calculated to relatively high accuracy. Cluster peculiar velocities therefore have great potential to provide constraints on cosmological parameters. Peculiar velocities for the template clusters calculated from combining their TF relations are shown in Table \ref{clusteroffsets}. In that table, (as well as in Table 1) clusters used to define the rest frame are marked with a * next to their CMB velocity. Three sets of peculiar velocities are reported, for both the \ins~ and \inp~ samples. $v_{\rm pec}(1)$ refers to offsets when no incompleteness bias corrections have been applied. $v_{\rm pec}(2)$ refers to the offsets when the simple additive version of the morphological correction is applied (along with all other bias corrections) and $v_{\rm pec}(3)$ refers to the offsets calculated when the width dependent morphological correction is applied. The error on these offsets is taken to be the scatter in the individual cluster TF relation divided by the square root of the number of galaxies in the cluster. 

 As explained in \citet{Wa97}, the 1D peculiar velocity dispersion of clusters depends relatively simply on cosmological parameters, a relationship coming from its dependence on the initial power spectrum of fluctuations. In a flat universe ($\Omega_M + \Omega_\Lambda = 1$) the relation is given by
\be
\sigma_v = \frac{100 {\rm km s}^{-1}}{\sqrt{3}} \Omega^{0.6} \sigma_8 \sqrt{f(\Omega h)}
\ee
where $\sigma_8$ is as usual the $rms$ fluctuation on scales of 8 Mpc/$h$, $\Omega$ is the mass density of the universe, and $h = H_0/100$ \kmsMpc. In \citet{Wa97} the function $f(\Omega h)$ is well fit by $f = 12.5 (\Omega h)^{-1.08} + 49.4$, and the approximation $\sqrt{f} \sim 10$ over the range $\Omega h = $ 0.2--0.5 is used. The third year WMAP data release quotes $\Omega h^2 = 0.134\pm0.006$, $h=0.73\pm0.03$ and $\sigma_8 = 0.84\pm0.06$ \citep{Sp06}, implying a value for the velocity dispersion of $\sigma = 239\pm23$ \kms. 

\begin{figure}
\epsscale{1}
\plotone{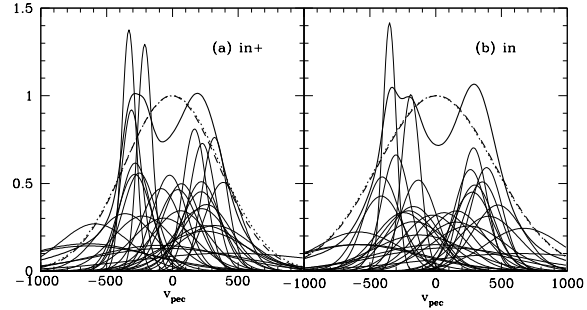}
\caption{Line of sight peculiar velocities in the cluster reference frame all clusters. Peculiar velocities are represented as equal area Gaussians, with the width being set by the error on the measurement. Panel (a) shows the results using the \inp ~sample for each cluster; Panel (b) shows the \ins~ sample. The heavy solid line in both panels is an arbitrarily scaled sum of all the Gaussians representing the individual cluster peculiar velocity measurements. The dashed and dotted lines are fits to the sum, the dotted line having the mean fixed to zero. When the mean is allowed to vary it is not significantly different from zero.  
\label{clustervpec}}
\end{figure}

 Figure \ref{clustervpec} shows our measured cluster peculiar velocity dispersion for the {\bf in+} and \ins~ samples (here the peculiar velocities are calculated from the mean TF offset after {\bf all} bias corrections to the TF data have been applied). The peculiar velocities of the clusters are represented as Gaussians of equal area, with the width being set by the error on the measurement. An arbitrarilly scaled sum of all these Gaussians is shown as the heavy solid line. A Gaussian fitted with zero mean to the entire {\bf in+} sample gives $\sigma = 365\pm34$ \kms, (or $\sigma=440\pm40$ \kms~ in the \ins ~sample). Here the errors are calculated using a Monte Carlo simulation in which 1000 realizations of the cluster peculiar velocities (drawn from the normal distribution shown individually for each cluster in Figure \ref{clustervpec}) is made. This measurement is biased larger than the intrinsic cluster peculiar velocity dispersion because of broadening due to the error on the individual cluster peculiar velocity measurements. A simplistic correction for this broadening, (subtracting the mean peculiar velocity error in quadrature) gives an intrinsic cosmic cluster velocity dispersion of $\sigma = 287\pm34$ \kms ~and $365\pm40$ \kms ~for the \inp ~and \ins ~samples respectively. Correcting for error broadening slightly more carefully using another Monte Carlo simulation gives $\sigma = 298 \pm 34$\kms for \inp ~and $\sigma = 368\pm 40$\kms for \ins.
 
 Our measurements of the cosmic cluster peculiar velocity dispersion are entirely consistent with those calculated by \citet{G98b} of 270$\pm54$ \kms ~and 277$\pm63$ \kms ~for their {\bf in+} and {\bf in} samples respectively. These values are also in agreement with other measures using the SCI data \citep{BO96,Wa97}, which all favor $\sigma_{1d} \sim 250$ \kms ~once values have been corrected for error broadening. 

 This measurement of the cosmic cluster peculiar velocity dispersion is also consistent with the WMAP estimate, providing yet another independent verification of the best fit cosmological model. We can use the approximation $\sqrt{f} \sim 10$ in Equation 7 to calculate a value $\Omega^{0.6} \sigma_8 = 0.52\pm0.06$ from this TF measurement of the cluster peculiar velocity dispersion alone (where the error here only represents the measurement error on the cluster peculiar velocity dispersion and does not account for any bias introduced by the assumption). Alternatively we can use WMAP information on $\Omega h^2=0.134\pm0.006$ (coming from the relative amplitude of the power spectrum peaks) along with our measurement of $h=0.74\pm0.02$ (see Section 8) from a combination of TF and Cepheid data to calculate $\sqrt(f)=11.34\pm0.59$, implying $\Omega^{0.6} \sigma_8 = 0.46\pm0.06$.

  These cluster peculiar velocities also provide tantalizing hints about the local flow field. It is interesting to note that the two clusters, Hydra and A3574 at $cz>4000$ \kms~ in the ``Great Attractor'' region both have slightly negative peculiar velocities (A3574 has zero peculiar velocity to within $1\sigma$), while the clusters at $cz<3500$ \kms~ in that direction have positive peculiar velocities. This is presumably the signature of infall onto the ``Great Attractor'' and could be interpreted as a detection of backside infall into that region. Infall onto the Pisces-Perseus supercluster is also hinted at. Abell 262 and NGC 383, both at $D \sim 65$ Mpc/$h_{75}$, have very marginal negative peculiar velocities in the CMB frame (N383 has $v_{\rm pec}=0$\kms ~to well within $1\sigma$) and are presumably falling into the supercluster from behind, while NGC 507 at $D \sim 60$ Mpc/$h_{75}$ in the same direction has a positive peculiar velocity indicating front-side infall. Also of note, the pair of clusters A2806 and A2877 have opposite peculiar velocities implying very similar real space distance, and presumably showing their motion towards each other and indicating a future cluster merger.  In the direction of the constellation Pavo, Pavo I (or the Pavo group) at $D\sim$50 Mpc appears to be in the foreground of Pavo II (or the Pavo cluster) at $D\sim 60$ Mpc and moving towards it.

\subsection{The Global TF Template}

\begin{figure} 
\epsscale{1}
\plotone{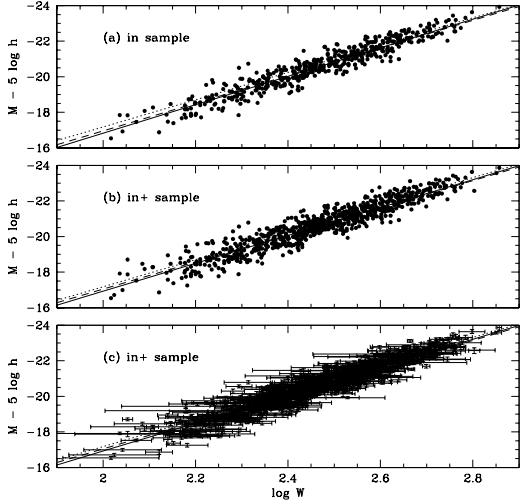}
\caption{Global TF relation for (a) {\bf in} sample, (b) {\bf in+} sample, (c) {\bf in+} sample plotted with measurement errors. The direct and bivariate fits to the respective samples are overlaid as dashed and solid lines respectively, while the G97b bivariate fit to their {\bf in+} sample is plotted as the dotted line.
\label{globalTF}}
\end{figure}

 The global template TF relation is shown in Figure \ref{globalTF} for both the {\bf in} and {\bf in+} samples. We also show the {\bf in+} sample plotted with error bars. A summary of linear fits (direct, inverse and bivariate) to various subsets of the full sample is shown in Table \ref{globalfits}. Quadratic fits do not improve the scatter around the relationship significantly and therefore are not reported. The first two lines in Table 4 give the TF fits for the \ins~ and \inp~ samples when no incompleteness bias (see Section 3.2) is applied. The direct and bivariate fits in these lines ought to have shallower slopes and brighter zeropoints than the bias corrected relation, which is observed. Next we show TF fits for incompleteness bias corrected magnitudes. The first set in Table 4 show the fits for different types of spiral galaxies. These fits are discussed in Section 3.5 and used to argue for a morphological type bias correction. Next we show fits using a plain additive morphological correction (as used in G97b, and discussed in Section 3.5) to account for the fact that at a given rotational width earlier type spirals are dimmer than later types. We show fits for \ins~ and \inp, and also for \inp ~divided into Scs and types earlier than Sc. Here it is clear that the slope for the earlier type spirals is shallower than the Sc slope (as previously discussed in Section 3.5), suggesting that a type dependent morphological is needed. This correction is applied to the magnitudes used in the fits shown in the last part of Table 4. It should be noted here that the addition of this width dependent morphological correction is in effect causing us to fit {\it two} TF relation to each of these sets of galaxies, one for the Scs and a second for galaxies earlier than Sc. Here we show fits for the \ins~ and \inp~ samples, but also the \inp~ sample with a $2.5\sigma$ clip applied to reduce the impact of outliers; for the \inp~ sample with inclination cuts applied since less inclined galaxies have larger errors on inclination corrections to the widths, and the most inclined galaxies can have large errors introduced from the impact of internal extinction. Finally we show the fit divided into source of rotation width (as discussed in Section 2.2.2), to show that there are no significant differences once the ORC correction derived in \citet{C06} is applied to correct ORC widths to a HI width scale. 

 The zeropoint derived for the TF relation from the ``basket of clusters'' method could have a systematic offset from the real value if there is a net motion of the cluster reference frame with respect to the CMB. We can estimate this offset under the assumption that the peculiar velocities of clusters are characterized by an $rms$ 1D velocity dispersion $\sigma_v$. The mean velocity of $N$ randomly chosen clusters with uncorrelated velocities is then $\langle v_{\rm pec}\rangle \sim \sigma_v N^{-1/2}$. In practice all 21 of our reference frame clusters do not constitute uncorrelated measurements of the peculiar velocity field, which is known to show coherent structures over large scales. A conservative estimate suggests we have $\sim 12$ independent measurements of the velocity field. We measured $\sigma_v = 298 \pm 34$ \kms ~in the preceding section, implying $\langle v_{\rm pec}\rangle \sim 86$ \kms ~for the set of reference clusters (assuming there are no coherent flows on the scale of the entire sample volume). The mean redshift of these clusters is $\langle cz \rangle = 6750 $\kms. The likely magnitude offset, $|\Delta M| \sim 2.17 \langle v_{\rm pec}\rangle \langle cz \rangle ^{-1}$, is therefore  $|\Delta M| \sim 0.03$ mag. This value should be combined in quadrature with the statistical error on the zeropoint of $\sim 0.015$ mag for the \inp ~sample (0.02 mag for \ins) to provide the final zeropoint accuracy of 0.03 mag for the \inp sample (or 0.04 mag for \ins).

\begin{deluxetable*}{llrrrrrr}
\tablecolumns{8} 
\tablewidth{0pc} 
\tablecaption{Fit offsets for individual cluster samples \label{clusteroffsets}} 
\tablehead{ 
\colhead{Sample}    &  \colhead{$v_{\rm CMB}$} & \multicolumn{2}{c}{$v_{\rm pec}(1)$ (\kms)} &\multicolumn{2}{c}{$v_{\rm pec}(2)$ (\kms)} &\multicolumn{2}{c}{ $v_{\rm pec}(3)$ (\kms)}\\
\colhead{} & \colhead{(\kms)}   & \colhead{\ins}    & \colhead{\inp} 
& \colhead{\ins}    & \colhead{\inp} & \colhead{\ins}    & \colhead{\inp} }
\startdata 
 N383& 4865* &  -24$\pm$ 227 & -115$\pm$ 254 &   -5$\pm$ 230 &  -58$\pm$ 244 &  -25$\pm$ 226 &  -76$\pm$ 239 \\
 N507& 4808* &  393$\pm$ 243 &  236$\pm$ 199 &  351$\pm$ 235 &  208$\pm$ 191 &  413$\pm$ 231 &  253$\pm$ 190 \\
  A262& 4665* & -328$\pm$ 212 & -151$\pm$ 157 & -248$\pm$ 225 &  -67$\pm$ 157 & -240$\pm$ 216 &  -82$\pm$ 153 \\
   A397& 9594* &  629$\pm$ 698 &  443$\pm$ 687 &  512$\pm$ 745 &  313$\pm$ 727 &  622$\pm$ 723 &  423$\pm$ 705 \\
  A400& 6934* &  196$\pm$ 217 &  267$\pm$ 141 &  137$\pm$ 208 &  225$\pm$ 141 &  150$\pm$ 210 &  229$\pm$ 141 \\
   A569& 6011* & -438$\pm$ 227 & -436$\pm$ 249 & -155$\pm$ 200 & -193$\pm$ 228 & -157$\pm$ 195 & -215$\pm$ 229 \\
   A634& 7922* & -773$\pm$ 459 & -831$\pm$ 448 & -587$\pm$ 487 & -664$\pm$ 479 & -577$\pm$ 475 & -663$\pm$ 462 \\
 Cancer& 4939* &   23$\pm$ 264 &    8$\pm$ 146 &  203$\pm$ 256 &   95$\pm$ 142 &  153$\pm$ 258 &   77$\pm$ 144 \\
   A779& 7211* & -599$\pm$ 302 & -628$\pm$ 255 & -616$\pm$ 317 & -603$\pm$ 264 & -582$\pm$ 323 & -600$\pm$ 267 \\
  A1314& 9970* & -107$\pm$ 514 & -189$\pm$ 520 &  332$\pm$ 522 &  230$\pm$ 529 &   84$\pm$ 544 &  -33$\pm$ 556 \\
  A1367& 6735* &  192$\pm$ 212 &  187$\pm$ 206 &   32$\pm$ 221 &   27$\pm$ 213 &   71$\pm$ 216 &   66$\pm$ 209 \\
 Ursa Major& 1101 & -392$\pm$ ~53 & -378$\pm$ ~54 & -352$\pm$  ~53 & -338$\pm$  ~54 & -348$\pm$  ~51 & -333$\pm$ ~52 \\
  Coma& 7185* &  619$\pm$ 165 &  599$\pm$ 155 &  323$\pm$ 142 &  358$\pm$ 143 &  361$\pm$ 142 &  398$\pm$ 142 \\
  A2199/7& 8996* &  -82$\pm$ 344 &  441$\pm$ 317 & -407$\pm$ 378 &  145$\pm$ 334 & -146$\pm$ 350 &  251$\pm$ 316 \\
 Pegasus& 3519 & -317$\pm$ 135 & -450$\pm$ 131 & -164$\pm$ 134 & -314$\pm$ 131 & -131$\pm$ 139 & -290$\pm$ 130 \\
  A2634& 8895* &  820$\pm$ 298 &  540$\pm$ 269 &  652$\pm$ 284 &  257$\pm$ 280 &  660$\pm$ 295 &  318$\pm$ 277 \\
\tableline
  A2806& 7867* &  340$\pm$ 273 &  325$\pm$ 270 &  395$\pm$ 312 &  365$\pm$ 309 &  266$\pm$ 280 &  236$\pm$ 279 \\
  A2877& 6974* & -309$\pm$ 379 & -339$\pm$ 390 & -361$\pm$ 462 & -407$\pm$ 481 & -596$\pm$ 479 & -647$\pm$ 502 \\
   A194& 5036* & -295$\pm$ 281 & -471$\pm$ 236 & -141$\pm$ 258 & -352$\pm$ 229 & -168$\pm$ 256 & -348$\pm$ 220 \\
 Eridanus& 1536 & -378$\pm$ 113 & -378$\pm$  ~80 & -307$\pm$ 108 & -316$\pm$  ~78 & -306$\pm$ 109 & -310$\pm$  ~78 \\
 Fornax& 1321 & -226$\pm$  ~72 & -260$\pm$  ~59 & -167$\pm$  ~74 & -202$\pm$ ~57 & -185$\pm$  ~72 & -209$\pm$  ~56 \\
   A496& 9809* &  -84$\pm$ 497 & -141$\pm$ 503 &   72$\pm$ 501 &   -6$\pm$ 512 &   15$\pm$ 479 &  -71$\pm$ 490 \\
 Antlia& 3120 &  290$\pm$ 105 &  203$\pm$  ~84 &  336$\pm$  ~99 &  205$\pm$  ~89 &  292$\pm$ 102 &  173$\pm$  ~89 \\
  Hydra& 4075* & -488$\pm$ 157 & -284$\pm$ 127 & -416$\pm$ 182 & -221$\pm$ 137 & -422$\pm$ 169 & -246$\pm$ 129 \\
 NGC3557& 3318 &  227$\pm$ 167 &  229$\pm$ 199 &  308$\pm$ 153 &  191$\pm$ 210 &  281$\pm$ 160 &  217$\pm$ 202 \\
  Cen30& 3322 &  216$\pm$ 135 &  278$\pm$ 100 &  249$\pm$ 126 &  331$\pm$  ~96 &  260$\pm$ 124 &  324$\pm$  ~94 \\
 ESO 508& 3196 &  411$\pm$ 129 &  232$\pm$ 102 &  396$\pm$ 113 &  254$\pm$  ~96 &  382$\pm$ 122 &  230$\pm$  ~99 \\
  A3574& 4817* & -190$\pm$ 209 &  -27$\pm$ 128 & -191$\pm$ 214 &  -20$\pm$ 134 & -199$\pm$ 210 &  -26$\pm$ 132 \\
 Pavo II& 4444* &  302$\pm$ 179 &   11$\pm$ 164 &  314$\pm$ 162 &   10$\pm$ 160 &  304$\pm$ 147 &   34$\pm$ 155 \\
 Pavo I& 4055 &  393$\pm$ 202 &  173$\pm$ 164 &  489$\pm$ 186 &  191$\pm$ 163 &  473$\pm$ 191 &  212$\pm$ 159 \\
 MDL59& 2317 & -535$\pm$ 135 & -347$\pm$ 128 & -411$\pm$ 137 & -303$\pm$ 113 & -410$\pm$ 134 & -277$\pm$ 117
\enddata
\tablenotetext{*}{Cluster used to define the rest frame for peculiar velocities in Section 5.1.}
\end{deluxetable*}

\begin{deluxetable*}{lccccccccccc}
\tablecolumns{12} 
\tablewidth{0pc} 
\tabletypesize{\scriptsize}
\tablecaption{Linear TF parameters for global samples.\label{globalfits}}
\tablehead{ 
\colhead{Sample} & \colhead{$N$} & \colhead{$a_{dir}$}    & \colhead{$b_{dir}$} & 
\colhead{$a_{inv}$} & \colhead{ $b_{inv}$}   & \colhead{$a_{bi}$}    & \colhead{$b_{bi}$} & 
\colhead{$\epsilon_a$}    & \colhead{$\epsilon_b$}   & \colhead{$\sigma_{bi}$}    & \colhead{$\sigma_{abs}$}} 
\startdata
\cutinhead{\bf No incompleteness correction}
{\bf in} &      486& -20.906&  -7.287& -20.883&  -8.294& -20.892&  -7.565&   0.019&   0.134&   0.384&   0.301 \\
{\bf in+} &      807& -20.915&  -7.190& -20.890&  -8.341& -20.901&  -7.442&   0.015&   0.105&   0.411&   0.321 \\
\cutinhead{\bf Incompleteness correction as described in Section 3.2}
\cutinhead{No morphological correction}
{\bf in+}, Sa    &   61& -20.826&  -5.199& -20.945&  -7.819& -20.774&  -5.516&   0.073&   0.403&   0.399&   0.317\\
{\bf in+}, Sb    &   281& -20.742&  -6.805& -20.847&  -8.506& -20.725&  -7.072&   0.025&   0.172&   0.379&   0.300\\
{\bf in+}, Sc    &   465& -20.855&  -7.685& -20.956&  -8.726& -20.853&  -7.870&   0.020&   0.145&   0.417&   0.328\\
\cutinhead{Plain additive morphological correction}
{\bf in} &     486& -20.820&  -7.318& -20.844&  -8.031& -20.808&  -7.561&   0.019&   0.134&   0.388&   0.303\\
{\bf in+} &     807& -20.833&  -7.192& -20.853&  -8.087& -20.823&  -7.425&   0.015&   0.106&   0.413&   0.324\\
{\bf in+}. Sa/Sb &   342& -20.873&  -6.606& -20.851&  -8.445& -20.845&  -6.919&   0.023&   0.154&   0.392&   0.310\\
{\bf in+}, Sc    &   465& -20.855&  -7.685& -20.956&  -8.726& -20.853&  -7.870&   0.020&   0.145&   0.417&   0.328 \\
\cutinhead{Width dependent morphological correction}
{\bf in} &      486& -20.841&  -7.765& -20.866&  -8.409& -20.829&  -8.006&   0.020&   0.137&   0.382&   0.299\\
{\bf in+} &    807& -20.859&  -7.625& -20.881&  -8.435& -20.849&  -7.845&   0.015&   0.103&   0.407&   0.320\\
{\bf in+} 2.5$\sigma$ clip  &794& -20.857&  -7.715& -20.898&  -8.455& -20.845&  -7.920&   0.015&   0.098&   0.384&   0.307\\
{\bf in+} $i>45$\arcdeg &  745& -20.855&  -7.710& -20.902&  -8.485& -20.847&  -7.868&   0.015&   0.102&   0.401&   0.315\\
{\bf in+} 60\arcdeg $< i < 80$\arcdeg & 411& -20.876&  -7.711& -20.922&  -8.476& -20.873&  -7.842&   0.020&   0.136&   0.393&   0.312\\
{\bf in+}, HI widths only  &577& -20.886&  -7.578& -20.920&  -8.481& -20.871&  -7.773&   0.017&   0.111&   0.412&   0.323\\
{\bf in+}, ORC widths only &230& -20.793&  -7.654& -20.840&  -8.364& -20.777&  -8.026&   0.030&   0.218&   0.371&   0.301
\enddata
\end{deluxetable*}

\subsection{Comparison with SCI TF Relation}
As discussed above, the sample here is based on the cluster sample SCI presented in \citet{G97a} to which a global TF relation is fit in G97b, but includes a significant amount of additional data, including 9 extra clusters (for which previously there was not sufficient numbers of galaxies with TF measurements to include) totaling $\sim 300$ new galaxies (or 30\% of the sample). The methods we have used to correct for biases, and to fit the TF relation are very similar to those discussed in G97b, so a direct comparison of the global TF relation should be made. The template sample here differs from the SCI sample mostly by the addition of small diameter and faint apparent magnitude galaxies, so for galaxies in a cluster all at basically the same distance this will preferentially add galaxies at the low mass end of the TF relation, adding extra leverage to the measurement of the slope, and increasing the overall scatter. As discussed earlier, the new instrumental corrections applied to the HI widths (which make up about 60\% of the velocity widths used) should also shallow the TF slope slightly relative to what would be found with the old corrections, and we also apply a correction to put ORC widths onto the same scale as the HI widths, which slightly shallows the slope and dims the zeropoint for these galaxies. There is also a significant difference in the distribution of different morphological types in the samples. SCI was designed to be heavily dominated by Sc galaxies, with 63\% of the template sample being Sc or later.  In this template sample 58\% of the galaxies are Sc or later, meaning that more than 50\% of the new galaxies are earlier type spirals. Here we apply a width dependent morphological type correction to the data since with the improved statistics the difference in slope between morphological types appears significant. With this correction our final TF relation fit to the \inp ~ sample is 
\be
M - 5 \log h = -20.85(2) -7.85(10)(\log W - 2.5), 
\ee
using the plain additive correction for morphological type as in G97b we find (for the same galaxies) $M - 5 \log h = -20.82(2) -7.43(11)(\log W - 2.5)$. The values of these slopes bracket the G97b result of $M - 5 \log h = -21.01(2) -7.68(10)(\log W - 2.5)$, which is what should be expected for a sample which has more early type spirals than the SCI sample. The zeropoints of the G97b TF relation and our favored fit differ by an amount $\Delta M = 0.16\pm0.03$, \ie ~ they are different at the $5\sigma$ level, such that the G97b value is $0.16\pm0.03$ mag brighter.  Part of this difference can be accounted for by the use of DIRBE \citep{SFD98} corrections for Galactic extinction, which are larger than those used in G97b (from \citealt{BH78}) by a mean value of 0.05 mag over lines of sight towards the template clusters, but this still leaves a difference of $\Delta M = 0.11\pm0.03$ mag. This can probably be explained by the use of the ORC correction from Section 2.2.2 (which was not available for ORC widths in the G97b template). This correction should dim the magnitude of the zeropoint of the whole sample slightly (because it increases the widths of galaxies with ORCs slightly while they remain at the same magnitude). Without applying it we find both a slightly larger cluster velocity dispersion (as discussed in Section 2.2 part of the difference between the ORC and HI widths can be absorbed into the cluster peculiar velocity dispersion because most clusters are biased towards the use of one method for measuring widths) and a zeropoint of $\sim$ -20.9 mag. The difference between this zeropoint and the G97b value (after adjusting for the DIRBE dust correction) is $\sim 0.1\pm0.03$ mag, now only a $3\sigma$ difference, although it should be noted here that since the samples are not independent, the significance of this agreement is not the same as for totally independent studies.

\section{The Scatter in the TF Relation}
\begin{figure} 
\epsscale{1}
\plotone{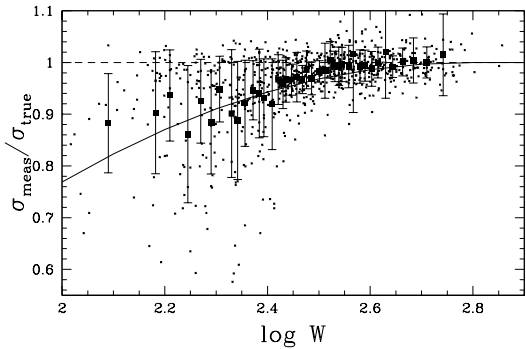}
\caption{Our calculation of the underestimate of the scatter in the TF relation due to the incompleteness bias for each \inp~ cluster. The bias for each galaxy was estimated from 1000 realizations of the individual \inp~ cluster samples with incompleteness characteristics derived in Section \ref{cbias}. The large squares show the mean bias in bins of $\sim$20 galaxies. The solid line is a parametric fit to the bias which we use to correct the measured scatter.
\label{scatterbias}}
\end{figure}

The intrinsic scatter in the TF relation carries as much (if not more) information for models of galaxy formation as the slope and intercept, and a proper understanding of the scatter is also important in deriving errors on distances (or peculiar velocities) from TF. As discussed in Section \ref{cbias}, the observed scatter in the TF relation of any sample with an implied or actual magnitude limit will be an underestimate of the true amount. Magnitude incompleteness in the sample means that galaxies which are intrinsically dimmer than the relation are preferentially removed when they are near the magnitude limit. This has the effect of shallowing the observed slope, brightening the observed zeropoint and decreasing the observed scatter (especially at the small width end of the relation). We model the bias, by making 1000 realizations of the global cluster sample, using the completeness for individual cluster samples as calculated in Section \ref{cbias}. The result for all galaxies in the \inp ~sample is shown in Figure \ref{scatterbias}, which also shows the average bias in bins of 20 galaxies, and a parametric fit which we use to correct the measured scatter. There is a wide range in the amount of bias at small $\log W$ because of the wide range of completeness at low absolute magnitudes found in the individual cluster \inp~ samples. Above $\log W =2.8$ we assume the bias to be negligible, and the parametric fit is fixed at that point. The bias correction for $\log W < 2.8$ is
\be
\frac{\sigma_{\rm measured}}{\sigma_{\rm true}} = 1.0 - 0.0044 (\log W - 2.8) - 0.367 (\log W - 2.8)^2.
\ee

\begin{figure} 
\epsscale{1}
\plotone{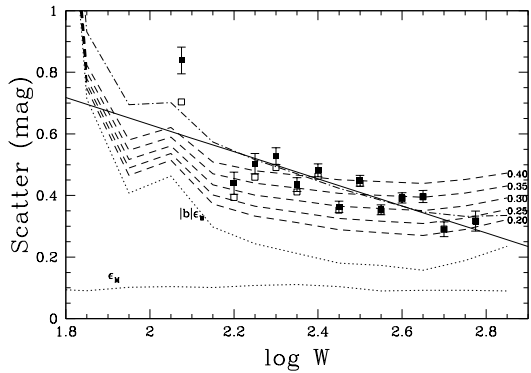}
\caption{Total observed scatter in the TF relation fit to the {\bf in+} sample. The open squares show the observed standard deviation. The filled squares have been corrected for incompleteness bias as shown the parametric fit in Figure \ref{scatterbias}. The solid line shows a linear fit to this bias corrected standard deviation of $\epsilon_{\rm obs} = 0.41 - 0.44 (\log W - 2.5)$. Also plotted is the error budget for all galaxies in the SFI++ dataset as a function of rotation width. The dotted lines show errors associated with the photometry and width measurement (multiplied by the TF slope to be expressed in magnitudes). The sum in quadrature of these two approximated the total measurement error. The dashed lines show a sum in quadrature of this total error with fixed scatters of 0.2, 0.25, 0.3, 0.35 and 0.4 mag respectively. The dot-dashed line shows a sum in quadrature of the total scatter with a width dependent intrinsic scatter of $\epsilon_{\rm int} = 0.35 - 0.37 (\log W - 2.5)$.
\label{scatterTF}}
\end{figure}

 The measured and bias corrected scatter (standard deviation) is shown by the open and solid square points in Figure \ref{scatterTF}. A linear fit to the bias corrected points gives $\epsilon_{\rm obs} = 0.41 - 0.44 (\log W - 2.5)$ (shown by the straight solid line in the figure). This total scatter is somewhat larger than $\epsilon_{\rm obs} = 0.32 - 0.325 (\log W - 2.5)$ found by G97b for the SFI template sample, but has a similar dependence on $\log W$. The source of the additional scatter must be related to the difference in the SFI and SFI++ samples in terms of both the distribution of galaxy properties and changes in the way measured values are calculated. The template sample of SFI++ adds to the SFI mostly small diameter galaxies, and also more early type spirals than in SFI which was designed to be dominated by Sc galaxies. These small galaxies are harder to assign proper morphological types (adding scatter in the morphological correction), and harder to measure inclinations for, adding a significant amount of scatter to the widths. The SFI++ template sample should therefore both have higher measurement errors on average and possibly also higher intrinsic scatter. We use the entire SFI++ sample (both template and field galaxies) to estimate the total measurement error at a given rotation width for SFI++ galaxies. These values are shown by the dotted lines in Figure \ref{scatterTF}. The lower line at $\sigma \sim 0.1$ shows the measurement error on the total magnitudes which at all widths is the least important source of error, and is similar to the value for SFI galaxies. The middle line shows a value of $7.85 \sigma_W$ which is the error on the measured rotation widths expressed in magnitudes. This quantity, particularly at small widths appears larger than that found for SFI galaxies. Adding these two contributions in quadrature approximates the total measurement error (this neglects the covariances between the two values which arise from the inclination corrections) which as expected is slightly larger than the total measurement error found for SFI. 

 The dashed lines in Figure \ref{scatterTF} show the total measurement error for SFI++ galaxies with fixed scatters of 0.2, 0.25, 0.3, 0.35 and 0.4 mag added in quadrature. As is obvious this does not provide a good fit to the measured scatter. At the low width end a much larger intrinsic scatter is needed than at the high width end even after the larger measurement errors have been accounted for. We therefore fit for an intrinsic scatter which has a width dependence and find $\epsilon_{\rm int} = 0.35 - 0.37 (\log W - 2.5)$. This fit is shown by the dot-dashed line in Figure \ref{scatterTF}, and provides a better fit to the observed error than the straight line fit. This value for the intrinsic scatter is again larger than that found in G97b, who measure $\epsilon_{\rm int} = 0.26 - 0.28 (\log W - 2.5)$ for the SFI cluster sample. Even accounting for the increased measurement errors in SFI++ there still appears to be a larger intrinsic scatter, presumably related to the larger variety of morphological types and perhaps the addition of higher redshift clusters into the template sample.

\section{A Cepheid Calibration of the TF zeropoint and $H_0$.}
 In this section we use the classic ``distance ladder'' approach to provide an independent calibration of the TF zeropoint. The ``basket of clusters'' method used above has several advantages over this classic approach, the primary being that it is independent of the Hubble constant. Even in today's era of ``precision'' cosmology, the error on the Hubble constant will significantly increase the error on the TF peculiar velocities measured using a template calibrated with the distance ladder method. Peculiar velocities from the ``basket of clusters'' method are independent of $H_0$. A further disadvantage of the distance ladder method is the still small numbers of calibrators. In the entire SFI++ sample of $\sim 5000$ galaxies, only 17 have distances measured from their Cepheid variables (data shown in Table \ref{cepheids}). These galaxies tend to lie at the high width end of the relation which could introduce a bias on the zeropoint. They also tend to have relatively low inclinations and thus the error on the determination of their actual rotation speeds is large. However, it is possible that there is a net motion of the reference cluster sample (see Section \ref{globalfit}) in the CMB frame which would bias the zeropoint measured with that method. The reference sample includes clusters with recessional velocities between 4000--10000 km/s, with a mean value of 6750 \kms. These clusters are distributed fairly evenly over the sky, but a small anisotropy in the distribution combined with the possibility of relatively large bulk flow towards the Shapley supercluster at $cz \sim 12000$ \kms, could introduce a bias. 

 It therefore seems prudent to check the ``basket of clusters'' calibration with an independent measure of the zeropoint from the distance ladder approach. Since there are no galaxies in the SFI++ sample with $W < 200$ \kms ~that have measured Cepheid distances, it is not appropriate to derive a slope from the calibrator sample, so we will use the slope derived in Section \ref{globalfit} and fit for the zeropoint alone. Under the assumption that the zeropoint calculated from the ``basket of clusters'' method does provide an unbiased measurement, this comparison alternatively provides an estimate of $H_0$. 

\begin{figure}
\epsscale{1}
\plotone{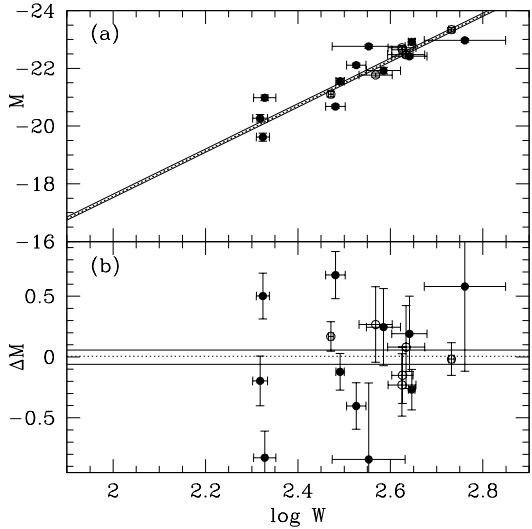}
\caption{Galaxies with measured distances from Cepheid variables plotted on the TF relation. Sc galaxies are shown as filled circles, earlier spirals as open circles. Absolute magnitudes have been corrected for morphological type dependence and the bivariate fit to the \inp ~ sample relation is shown as the two solid line (adjusted for $H_0 = 74\pm2$\kmsMpc). The bottom panel shows the residuals from this relation. Errors on the residuals are the sum in quadrature of the measurement errors on the magnitudes, the Cepheid distances (including the metallicity dependence) and the error on the widths multiplied by the slope of the template (this last error dominates).  The dotted line shows the error weighted mean residual.  \label{cephfig}}
\end{figure}

 The bulk of the Cepheid distances for SFI++ galaxies come from the HST Key Project (Freedman \etal ~2001; F01). Two additional distances come from Leonard \etal ~(2003; NGC 1637) and Newman \etal~ (1999; NGC 4603, this distance has been adjusted from that reported in the paper to be consistent with the newer Cepheid calibrations). We give in Table \ref{cepheids} the extinction corrected distance moduli reported in those papers. Cepheid distances are known to have a dependence on the metallicity of the host galaxy \citep{S04}. We use the \citet{S04} correction of $(m-M)_{\rm cor} = (m-M) - (0.24\pm 0.05) ($[O/H]$_{\rm gal}-$[O/H]$_{\rm LMC})$ with [O/H]$_{\rm LMC}$ = 8.50 dex, propagating the uncertainty in this correction through to the final Cepheid distance modulus uncertainty. The apparent magnitudes listed in Table \ref{cepheids} have been corrected for Galactic and internal extinction and the width-dependent morphological type correction from Section 3.5. All but two of these widths come from HI measurements. The two ORC widths for galaxies in the Fornax cluster are noted in Table \ref{cepheids} and small corrections from Equation \ref{ORCcorrection} have been applied (both of the ORCs are flat so the corrections are very small). Panel (a) of Figure \ref{cephfig} shows the galaxies with Cepheid distances on the Tully-Fisher relation, while their residuals from the bivariate fit to the \inp~ sample are shown in panel (b). While the TF relation from Section \ref{globalfit} is derived independent of the value of $H_0$, a value must be used to compare with the Cepheid calibrated zeropoint. In Figure \ref{cephfig} the best fit relation has been adjusted for $H_0 = 74\pm2$ \kmsMpc ~(\citealt{Sa06} from a combination of WMAP and 2dFRGS); the upper and lower limits on $H_0$ are shown by the two solid lines in both panels. The errors on the residuals for the individual galaxies show the sum in quadrature of the measurement error on the apparent magnitudes, the error on the Cepheid distance, and the error on the measured width multiplied by the slope of the TF relation. The error on the residuals for all these galaxies is dominated by the measurement error on the galaxy widths. 

\begin{deluxetable*}{lcclcclcll}
\tablecolumns{9} 
\tablewidth{0pc} 
\tabletypesize{\footnotesize}
\tablecaption{Galaxies in SFI++ with Cepheid Distances\label{cepheids}}
\tablehead{ 
\colhead{Galaxy} & \colhead{$v_{\rm CMB}$}   & \colhead{$(m-M)$}   & \colhead{[O/H]\tablenotemark{a}} & \colhead{$(m-M)$\tablenotemark{b}} 
& \colhead{ $m$} & \colhead{$\log W$}       & \colhead{Inc}   & \colhead{Type}    & \colhead{Cluster} \\
\colhead{} & \colhead{(\kms) }   & \colhead{(mags)} & \colhead{(dex)} & \colhead{(mags)}   & \colhead{mags} & 
\colhead{} & \colhead{deg}   &  \colhead{} & \colhead{}}
\startdata 
NGC 925 &   327& 29.80$\pm$0.04 & 8.55$\pm$0.15 & 29.81 & ~8.83$\pm$  0.10  &  2.328$\pm$ 0.024~  & 65.6 &  Sc & \\
NGC 1365          &  1542& 31.18$\pm$0.05 & 8.96$\pm$0.20 & 31.29 & ~8.32$\pm$  0.06  &  2.761$\pm$ 0.088\tablenotemark{c}  & 39.0 &  Sc & Fornax\\
NGC 1425          &  1400& 31.60$\pm$0.05 & 9.00$\pm$0.15 & 31.72 & ~9.61$\pm$  0.07  &  2.526$\pm$ 0.021\tablenotemark{c}  & 62.0 &  Sc & Fornax\\
NGC 1637          &   671& 30.23$\pm$0.07 & 9.08$\pm$0.15 & 30.37 & ~9.69$\pm$  0.06 &  2.481$\pm$ 0.021~  & 34.9 &  Sc &\\
NGC 2090          &   994& 30.29$\pm$0.04 & 8.80$\pm$0.15 & 30.36 & ~9.36$\pm$  0.10  &  2.471$\pm$ 0.005~  & 66.1 &  Sb & \\
NGC 2541          &   696& 30.25$\pm$0.05 & 8.50$\pm$0.15 & 30.25 & 10.63$\pm$  0.14  &  2.324$\pm$ 0.014~  & 60.9 &  Sc & \\
NGC 3198          &   877& 30.68$\pm$0.08 & 8.60$\pm$0.15 & 30.70 & ~9.15$\pm$  0.10  &  2.491$\pm$ 0.009~  & 71.2 &  Sc & \\
NGC 3319          &   979& 30.64$\pm$0.09 & 8.38$\pm$0.15 & 30.61 & 10.34$\pm$  0.13  &  2.318$\pm$ 0.016~  & 67.8 &  Sc &\\
NGC 3351 &  1125& 29.85$\pm$0.09 & 9.24$\pm$0.20 &30.03 & ~8.36$\pm$  0.07  &  2.568$\pm$ 0.036~  & 43.8 &  Sb & \\
NGC 3368 &  1235& 29.97$\pm$0.06 & 9.20$\pm$0.20 &30.14 & ~7.82$\pm$  0.12  &  2.626$\pm$ 0.023~  & 48.4 &  Sab & \\
NGC 4321 &  1897& 30.78$\pm$0.07 & 9.13$\pm$0.20 &30.93 & ~8.17$\pm$  0.06  &  2.553$\pm$ 0.079~  & 42.4 &  Sc & Virgo\\
NGC 4414          &   981& 31.10$\pm$0.05 & 9.20$\pm$0.15 & 31.27 & ~8.85$\pm$  0.05  &  2.641$\pm$ 0.038~  & 50.9 &  Sc & \\
NGC 4535          &  2295& 30.85$\pm$0.05 & 8.85$\pm$0.15 &30.93 & ~9.01$\pm$  0.11  &  2.585$\pm$ 0.037~  & 45.0 &  Sc & Virgo\\
NGC 4548 &   810& 30.88$\pm$0.05 & 9.34$\pm$0.15 &31.08& ~8.61$\pm$  0.11  &  2.634$\pm$ 0.040~  & 35.0 &  Sb & Virgo\\
NGC 4603          &  2877& 32.67$\pm$0.11 & 8.90$\pm$0.20\tablenotemark{d} &32.77& ~9.85$\pm$  0.10 &  2.646$\pm$ 0.007~  & 51.0 &  Sc & Cen 30\\
NGC 4725          &  1489& 30.38$\pm$0.06 & 8.92$\pm$0.15 &30.48& ~7.77$\pm$  0.07  &  2.625$\pm$ 0.030~  & 62.7 &  Sb & Virgo\\
NGC 7331          &   491& 30.81$\pm$0.09 & 8.67$\pm$0.15 &30.85& ~7.61$\pm$  0.08  &  2.732$\pm$ 0.006~  & 64.9 &  Sb & 
\enddata
\tablenotetext{a}{12+log(O/H)}
\tablenotetext{b}{Corrected for metallicity effects using the \citet{S04} correction}
\tablenotetext{c}{Widths from ORCs corrected to $W_{\rm HI}$ scale using Equation 2. Both galaxies have flat ORCs.}
\tablenotetext{d}{Estimated metallicity as in \citet{N99}}
\end{deluxetable*}

The error weighted mean of the residuals for these galaxies gives $M_{\rm Cepheid}-M_{\rm Clusters} = -0.01 \pm 0.05 \pm 0.06$. Here the first error quoted is the statistical error on the weighted mean, the second takes into account the 3\% error on $H_0$. This fit is shown by the dotted line in both panels of Figure \ref{cephfig}. It is indistinguishable from zero. The Cepheid calibrated TF relation (using the slope from the bivariate fit to the \inp ~ sample) is therefore
\be
M = -20.485 \pm 0.05 \pm 0.06 - (7.85 \pm 0.1) (\log W - 2.5),
\ee
(where again the first error is statistical and the second due to the uncertainty on $H_0$). This is to be compared to 
\be 
M - 5 \log h_{74} = -20.474 \pm 0.03 - (7.85 \pm 0.1) (\log W - 2.5)
\ee 
from the basket of clusters method. Here we note again that this template is independent of $H_0$, but has been adjusted to $H_0 = 74$ \kmsMpc ~ for ease of comparison. It is certainly reassuring that the two methods for calibrating the zeropoint agree so well. This agreement provides a strong limit on how much the cluster reference frame differs from the CMB rest frame. 

Under the assumption that the ``basket of clusters'' method provides an unbiased measurement of the TF zeropoint, we can use the combination of that TF relation and the independent Cepheid calibration of the zeropoint to provide an estimate of $H_0$. Equating the zeropoints from the two methods is equivalent to using the Cepheid calibrated TF relation to measure redshift-independent distances for the 31 clusters in our sample and taking the mean of the values of $H_0$ calculated from $v/D$ for each cluster. Since we account for the cluster peculiar velocities when combining their TF relations (see Section 5) here $v$ is already corrected for the peculiar motion of each cluster. By equating the two zeropoint we derive $H_0 = 74\pm2$ (random) \kmsMpc. This measurement has two possible sources of systematic error. The first is that the ``basket of clusters'' template may be measure a biased value for the TF zeropoint if there is a net motion of the reference frame. This point is discussed in more detail in Section 5.2, where we estimate at most a systematic offset of 0.03 mag which is included in the cluster TF zeropoint uncertainty (and therefore in the error already quoted for $H_0$). A larger source of possible systematic error comes from the Cepheid distance scale. The error comes both a combination of the uncertain metallicity dependence of Cepheid distances \citep{S04}, WFPC2 calibration systematics, and uncertainty on the distance to the LMC which sets the zeropoint for Cepheid distances. We account for the metallicity dependence of the Cepheid distance scale using the correction derive in \citet{S04}. Almost all galaxies in the Cepheid--SFI++ sample are more metal rich than the LMC, resulting in an overestimate of $H_0$ if the dependence is not accounted. If the metallicity dependence is not corrected for we measure $H_0=77\pm2$ \kms, implying that the maximum systematic error which could be introduced if this correction is wrong is $\sim \pm2$ \kms. The effect of the uncertainty on the WFPC2 calibration and the distance to the LMC are discussed in detail in \citet{S00} where a total systematic error on the Cepheid zeropoint of 0.16 mag is quoted. This adds a error of $\pm6$ \kms ~to our estimate of $H_0$, resulting in a final estimate of $H_0 = 74 \pm 2 \pm 6$ \kms. 

 Our measurement of $H_0$ is identical to the determination of $H_0 = 74\pm2$ \kmsMpc ~\citep{Sa06} from a combination of WMAP and 2dFRGS data, and the recent measurement of $H_0=74\pm3\pm6$\kmsMpc ~which uses a new Cepheid calibration from NGC 4258 in combination with SNIa distances \citep{M06}. Other measurements of $H_0$ using Cepheid calibrations of the TF relation include the HST Key Project value of $H_0=71\pm4\pm7$\kmsMpc ~\citep{S00}, and the SFI measurement of $H_0 = 69 \pm 2 \pm 6$ \kms ~(Giovanelli \etal ~1997a; note that neither of these determinations accounted for the metallicity dependence of the Cepheid zeropoint). We point out that the dominant source of error in our determination of the value of $H_0$ comes not from the TF relation, but from the uncertainty in the zeropoint calibration of the Cepheid relation, and ultimately from the uncertainty of our knowledge of the distance to the LMC. If this error were not present TF could measure $H_0$ to a comparable accuracy to the best ``precision cosmology'' available, and better than is possible from just WMAP alone which quotes $H_0 = 73 \pm 3$\kms ~and also requires an assumption that the universe is flat \citep{Sp06}. In fact, work on improving the Cepheid calibration is underway. \citet{M06} discuss the possibility of improving $H_0$ measurements based on the Cepheid scale to $\pm 5$\% in the near future, noting that soon there will be four galaxies with geometric distance measures (NGC 4258, LMC, M31, M33) which could be used to set the Cepheid zeropoint.

\section{Conclusions}

 In this paper we present the SFI++ sample of spiral galaxies. Galaxies in this sample have both I-band photometry and velocity widths (either from HI global profiles or optical rotation curves) suitable for use in the Tully-Fisher relation. From a subset of this sample consisting of 807 galaxies in the vicinity of 31 nearby clusters we rederive the I-band TF relation. This sample constitute by far the largest ever available to calibrate the TF relation. We argue that it provides evidence for a type-dependent TF slope that is steeper for later type spirals, in effect fitting for two TF relations -- one for the 465 Sc galaxies in the sample, the other for the 342 Sa/Sb galaxies. We measure the cluster peculiar velocity dispersion for the 31 cluster samples. This quantity is directly related to the shape of the initial power spectrum of density fluctuations and is used to estimate a value of $\Omega^{0.6}\sigma_8=0.52\pm0.06$. By comparing the TF template with the Cepheid distances to 17 galaxies we measure $H_0=74\pm2\pm6$ \kmsMpc; an accuracy comparable to the best precision cosmology. A summary of the numeric results is presented in Table \ref{conclusions}. A more detailed summary of the conclusions follows in this section. 

\begin{deluxetable*}{ll}
\tablecolumns{1} 
\tablewidth{0pc} 
\tablecaption{Summary of Numeric Results\label{conclusions}}
\tablehead{\colhead{Parameter} & \colhead{Values}}
\startdata 
Template relation (bivariate fit to \inp ~sample) & 
  $M - 5 \log h = -20.85 \pm 0.03 - (7.85 \pm 0.1)(\log W - 2.5)$ mag \\
 &  $L = 3.7 \times 10^{10} (v_{\rm max}/200 ~{\rm km~s}^{-1})^{3.1} L_\sun/h^2$\\
Template scatter  & 
$\epsilon_{\rm total} = 0.41 - 0.44 (\log W - 2.5)$ mag\\
& $\epsilon_{\rm intrinsic} = 0.35 - 0.37 (\log W - 2.5)$ mag\\
Cepheid calibrated zeropoint & 
$M = -20.49 \pm 0.05 \pm 0.06 - (7.85\pm0.1)(\log W - 2.5)$ mag\\
Hubble's constant & 
   $H_0=74\pm2\pm6$ \kmsMpc \\
Cluster peculiar velocity dispersion &
  $\sigma = 298\pm34$ \kms \\
&  $\Omega^{0.6} \sigma_8 =0.52\pm0.06$
\enddata
\end{deluxetable*}

 In Section 3.1 we introduced a subset of the SFI++ sample which consists of 807 galaxies in the vicinity of 31 nearby clusters (within $v_{\rm CMB}=$10,000 \kms). This sample is divided into an {\bf in} sample consisting of 483 {\it bona fide} cluster galaxies and an {\bf in+} sample which also includes galaxies considered to be peripheral members of the cluster. Cluster membership assignments are discussed further in Springob \etal ~(2007). 
 
 In Section \ref{bias}.2, we discuss various biases which modify the observed TF relation from the intrinsic relation within a cluster sample. We derive corrections for these biases which include morphological type bias and incompleteness bias due to the implicit magnitude limit of the sample. The magnitude of the incompleteness bias depends most strongly on the assumed scatter in the TF relation. 

 We find evidence to suggest that the slope of the TF relation gets shallower for early type spirals and construct a width-dependent morphological correction to reconstruct the relation for Sc galaxies. This is equivalent to fitting two separate TF relations to the sample - one to the major subset of 465 Sc galaxies, a second to the 342 galaxies in the sample with types earlier than Sc. The slope difference can be explained either by differing completeness characteristics for the different morphological types in the sample, or a real physical difference in the galaxies. We argue for the latter as the difference is shown to be largest at the high width end of the relation where the bias corrections are smallest. In the numerical models of \citet{M00} and \citet{NS00} this appears consistent with the idea that earlier type spirals have a smaller fraction of their mass in the disk at a given rotational velocity, and implies that as well as being dimmer than later types at a given rotational velocity they have less concentrated halos. 
 
 Any TF relation fit to a sample with an explicit or implicit magnitude limit will suffer from biases relating to incompleteness. The largest such bias is caused by it being possible to observe galaxies which are scattered brighter than the TF relation from near the magnitude limit, while those scattered dimmer will not make it into the sample. We calculate and correct for this bias, as well as those relating to the finite depth of the clusters. All bias corrections applied are summarized in Section 3.6.

 Individual TF relations are fit to each of the 31 clusters samples and presented in Section \ref{indivTF}. We search for environmental dependence of the TF relation by looking for correlations in global cluster properties with the slope, zeropoint and scatter. No correlations are found. There is also no dependence of TF residuals on projected cluster center difference for galaxies in the sample. We therefore argue that there is no evidence for environmental dependence of the TF relation. It is noted however that because of the morphology--density relation we are already implicitly correcting for an environmental dependence by applying morphological type corrections.
 
 The individual cluster samples are combined in Section \ref{globalfit} to create a global TF template. In order to do this, peculiar velocities are derived for the clusters and a cluster velocity dispersion of $\sigma = 298\pm 34$ \kms ~is measured. This measurement provides information on cosmological parameters as it depends on the initial spectrum of density fluctuations. The value is completely independent, yet within $2\sigma$ of that predicted from the best fit cosmological model to the third year WMAP data release which implies $\sigma = 239\pm23$ \kms \citep{Sp06}. With minimal assumptions about cosmology we use the measurement to estimate $\Omega^{0.6} \sigma_8 = 0.52\pm0.06$ from TF data alone.

Direct, inverse and bivariate linear fits to the combined {\bf in} and {\bf in+} samples (and subsets of them) are presented. Quadratic fits do not reduce the observed scatter and so are not reported. We favor the bivariate fit to the {\bf in+} sample, which gives a template TF relation of $M - 5 \log h = -20.85 - 7.85 (\log W - 2.5)$, or $L \propto v^{3.1}$. The statistical error on the zeropoint is $\sim 0.02$ mag, an additional 0.03 mag is estimated to account for the possible net motion of the cluster reference frame in the CMB. The statistical error on the slope is $\sim 0.10$ mag$^{-1}$. The TF relation obtained here will be applied to the rest of the galaxies in the SFI++ sample in future papers including Masters \etal ~(2007, in prep.) and Springob \etal ~(2007) to study the local peculiar velocity field.

The scatter in the TF relation is as important as the relation itself, both in models of galaxy formation and evolution and in deriving reliable distance errors.  We measure the total incompleteness bias corrected scatter from the template relation to be $\epsilon = 0.41 - 0.44 (\log W - 2.5)$. Once measurement errors have been accounted for this results in an intrinsic TF scatter of  $\epsilon = 0.35 - 0.37 (\log W - 2.5)$. This scatter corresponds to distance errors of 10\% for the largest width galaxies increasing to 26\% for the smallest widths.   

 As an independent measure of the zeropoint, that does not rely on the assumption that a subset of the most distant clusters are at rest with the CMB, we use 17 galaxies in the SFI++ that also have published distances from Cepheid variables. We find a negligible mean offset in the zeropoint of the TF relation for these galaxies relative to the ``basket of clusters" bivariate fit to the \inp~ sample when $H_0=74\pm2$ \kmsMpc ~is assumed, providing a reassuring check that the ``basket of clusters" method provides a reliable measure of the TF zeropoint. Turning this method around to estimate $H_0$ from a combination of the template and Cepheid distances we find $H_0=74\pm2\pm6$ \kmsMpc. The dominant source of error on this measurement comes not from TF directly, but rather from the uncertainty on the distance to the LMC (via the Cepheid calibration).  

 The TF relation still has a place in the era of ``precision cosmology'' to provide independent checks of the best fit cosmological model. It has the potential to measure some parameters (notably the cluster peculiar velocity dispersion and $H_0$) to an accuracy comparable to the best available. In addition the TF relation and its scatter provides valuable information about the formation of disk galaxies. However, the reliability of any conclusion drawn from TF depends critically on the availability of an unbiased template relation, such as the one provided here.

\begin{acknowledgements}
This work is based on a substantional amount of data taken at the Arecibo
Observatory. The Arecibo Observatory is part of the National Astronomy and
Ionosphere Center, which is operated by Cornell University under a
cooperative agreement with the National Science Foundation.
We wish to thank numerous collaborators and members of the Cornell Extragalactic Group without whose work the SFI++ sample would not have been possible. In particular we want to mention the many hours of observation and data reduction which went into assembling the SFI++ catalog. Barbara Catinella deserves a special mention especially for her work on matching HI and ORC widths. We also want to thank Lucas Macri especially for his comments on Section 7, and Michael Wood-Vasey for careful reading and comments which significantly improved Section 1. This work has been partially supported by NSF grants AST-9900695, AST-0307661 and AST-0307396 and was completed while KLM was a Harvard Postdoctoral Research Fellow supported by NSF grant AST-0406906. CMS was supported by the NRAO/GBT 03B-007 Graduate Student Support Grant and the NASA New York Space Grant while at Cornell and now holds a National Research Council Research Associateship at the Naval Research Laboratory. Basic research in astronomy at the Naval Research Laboratory is funded by the Office of Naval Research. 

\end{acknowledgements}

\end{document}